\pdfoutput=1
\documentclass[11pt,english]{article}
\usepackage[T1]{fontenc}
\usepackage[english]{babel}

\usepackage{mathptmx}
\usepackage{geometry}
\usepackage{calc}
\usepackage{setspace}
\usepackage{pdfpages}
\usepackage{amssymb,amsmath}
\usepackage{rotating}
\usepackage{chngcntr}
\usepackage{ragged2e}
\usepackage{pdflscape}
\usepackage{multirow}
\usepackage{float}
\usepackage{array}
\usepackage{appendix}
\usepackage{fancyhdr}
\usepackage{xcolor}
\usepackage{changepage}
\usepackage{comment}
\usepackage{bbm}
\usepackage{graphicx}
\usepackage{subcaption}
\usepackage{titlesec}
\usepackage{enumitem}
\usepackage{etoolbox}
\usepackage{adjustbox}
\usepackage[section]{placeins}
\usepackage{eso-pic}
\usepackage{listings}

\usepackage[font=large,figurename=FIGURE,tablename=TABLE]{caption}

\DeclareCaptionLabelFormat{panel}{Panel~#2}
\makeatletter
\renewcommand{\p@subfigure}{\thefigure.}
\makeatother

\DeclareCaptionStyle{topcap}{
  position=top,
  skip=0.5em,
  labelsep=none,
  justification=centering,
  format=plain,
  font={singlespacing}
}
\makeatletter
\newcommand{\fignotes}[1]{%
  \par\addvspace{0.5em}%
  {%
    \captionsetup{
      justification=justified,
      singlelinecheck=false,
      width=\linewidth,
      font=scriptsize,
      skip=0pt,
      aboveskip=0pt,
      belowskip=0pt,
      parskip=0pt,
      strut=off
    }%
    \caption*{\textit{Notes:}~#1}%
  }%
}

\makeatother

\makeatletter
\def\@maketitle{%
  \newpage
  \null
  \vspace*{-2em}%
  \begin{center}%
    \let\footnote\thanks
    {\setstretch{1}\LARGE \@title \par}%
    \vskip 1.5em%
    {\setstretch{1}\large
      \lineskip .5em%
      \begin{tabular}[t]{c}%
        \@author
      \end{tabular}\par}%
    \vskip 1em%
    {\large \@date}%
  \end{center}%
  \par
  \vskip 1.5em%
}

\newif\ifTB@skipnextpagenum
\TB@skipnextpagenumfalse
\pretocmd{\maketitle}{\global\TB@skipnextpagenumtrue}{}{}
\makeatother

\usepackage{natbib}

\usepackage{booktabs}
\usepackage{makecell}
\usepackage{longtable}
\usepackage{threeparttable}
\usepackage{threeparttablex}
\usepackage{datatool}

\usepackage{dcolumn}
\newcolumntype{d}[0]{D{.}{.}{5}}

\newcolumntype{L}[1]{>{\raggedright\let\newline\\\arraybackslash\hspace{0pt}}b{#1}}
\newcolumntype{C}[1]{>{\centering \let\newline\\\arraybackslash\hspace{0pt}}b{#1}}
\newcolumntype{R}[1]{>{\raggedleft\let\newline\\\arraybackslash\hspace{0pt}}b{#1}}

\makeatletter
\newcommand\BaseFontSizeTen{%
\renewcommand\normalsize{%
   \@setfontsize\normalsize\@xpt\@xiipt
   \abovedisplayskip 10\p@ \@plus2\p@ \@minus5\p@
   \abovedisplayshortskip \z@ \@plus3\p@
   \belowdisplayshortskip 6\p@ \@plus3\p@ \@minus3\p@
   \belowdisplayskip \abovedisplayskip
   \let\@listi\@listI}
\normalsize
\renewcommand\small{%
   \@setfontsize\small\@ixpt{11}%
   \abovedisplayskip 8.5\p@ \@plus3\p@ \@minus4\p@
   \abovedisplayshortskip \z@ \@plus2\p@
   \belowdisplayshortskip 4\p@ \@plus2\p@ \@minus2\p@
   \def\@listi{\leftmargin\leftmargini
               \topsep 4\p@ \@plus2\p@ \@minus2\p@
               \parsep 2\p@ \@plus\p@ \@minus\p@
               \itemsep \parsep}%
   \belowdisplayskip \abovedisplayskip
}
\renewcommand\footnotesize{%
   \@setfontsize\footnotesize\@viiipt{9.5}%
   \abovedisplayskip 6\p@ \@plus2\p@ \@minus4\p@
   \abovedisplayshortskip \z@ \@plus\p@
   \belowdisplayshortskip 3\p@ \@plus\p@ \@minus2\p@
   \def\@listi{\leftmargin\leftmargini
               \topsep 3\p@ \@plus\p@ \@minus\p@
               \parsep 2\p@ \@plus\p@ \@minus\p@
               \itemsep \parsep}%
   \belowdisplayskip \abovedisplayskip
}
\renewcommand\scriptsize{\@setfontsize\scriptsize\@viipt\@viiipt}
\renewcommand\tiny{\@setfontsize\tiny\@vpt\@vipt}
\renewcommand\large{\@setfontsize\large\@xiipt{14}}
\renewcommand\Large{\@setfontsize\Large\@xivpt{18}}
\renewcommand\LARGE{\@setfontsize\LARGE\@xviipt{22}}
\renewcommand\huge{\@setfontsize\huge\@xxpt{25}}
\renewcommand\Huge{\@setfontsize\Huge\@xxvpt{30}}
}
\makeatother

\setlist[enumerate,1]{label=(\roman*)}

\renewcommand{\thesection}{\Roman{section}}
\renewcommand{\thesubsection}{\thesection.\Alph{subsection}}

\renewcommand{\thefigure}{\arabic{figure}}

\makeatletter
\renewcommand{\p@subsection}{}%
\makeatother

\titleformat{\section}{\Large\bfseries}{\thesection}{0.5em}{}
\titleformat{\subsection}{\large\bfseries}{\thesubsection}{0.5em}{}

\makeatletter
\AtBeginDocument{%
  \setlength{\@fptop}{0pt}
  \setlength{\@fpbot}{0pt plus 1fil}
  \setlength{\@fpsep}{\dimexpr\textheight+1pt\relax}
}
\makeatother

\makeatletter
\newif\ifTB@inlandscape
\TB@inlandscapefalse

\newif\ifTB@hasfloats
\newcommand{\TB@maybe@clearpage}{%
  \par
  \TB@hasfloatsfalse
  \ifx\@deferlist\@empty\else\TB@hasfloatstrue\fi
  \ifx\@toplist\@empty\else\TB@hasfloatstrue\fi
  \ifx\@botlist\@empty\else\TB@hasfloatstrue\fi
  \ifx\@dbltoplist\@empty\else\TB@hasfloatstrue\fi
  \ifx\@dblbotlist\@empty\else\TB@hasfloatstrue\fi
  \ifdim\pagetotal>0pt
    \clearpage
  \else
    \ifTB@hasfloats\clearpage\fi
  \fi
}

\newlength{\TBHeaderPad}
\newlength{\TBBaseLeftOdd}
\newlength{\TBBaseLeftEven}
\AtBeginDocument{%
  \setlength{\TBBaseLeftOdd}{\dimexpr 1in+\hoffset+\oddsidemargin\relax}%
  \setlength{\TBBaseLeftEven}{\dimexpr 1in+\hoffset+\evensidemargin\relax}%
}

\newcommand{\TBSetHeaderPadToBase}{%
  \ifodd\value{page}%
    \setlength{\TBHeaderPad}{\TBBaseLeftOdd}%
  \else
    \setlength{\TBHeaderPad}{\TBBaseLeftEven}%
  \fi
}

\newcommand{\TBResetHeaderPad}{\setlength{\TBHeaderPad}{0pt}}

\newif\ifTBInAppendix
\TBInAppendixfalse
\AtBeginEnvironment{appendices}{\TBInAppendixtrue}
\AtEndEnvironment{appendices}{\TBInAppendixfalse}

\BeforeBeginEnvironment{landscape}{%
  \FloatBarrier
  \TB@maybe@clearpage
  \TB@inlandscapetrue
  \pagestyle{empty}% avoid rotated fancy headers in landscape
  \newgeometry{tmargin=0in,bmargin=0in,lmargin=1in,rmargin=1in}%
}

\AfterEndEnvironment{landscape}{%
  \clearpage
  \restoregeometry
  \TB@inlandscapefalse
  \pagestyle{fancy}%
}

\BeforeBeginEnvironment{table}{%
  \ifTB@inlandscape\else
    \FloatBarrier
    \TB@maybe@clearpage
    \newgeometry{tmargin=1in,bmargin=1in,lmargin=0in,rmargin=0in}%
    \ifTBInAppendix
      \setlength{\TBHeaderPad}{1in}%
    \else
      \TBSetHeaderPadToBase%
    \fi
  \fi
}

\AfterEndEnvironment{table}{%
  \ifTB@inlandscape\else
    \clearpage
    \restoregeometry
    \TBResetHeaderPad
  \fi
}

\BeforeBeginEnvironment{figure}{%
  \ifTB@inlandscape\else
    \TBResetHeaderPad
  \fi
}

\AfterEndEnvironment{figure}{%
  \ifTB@inlandscape\else
    \TBResetHeaderPad
  \fi
}

\def\fps@table{t}
\makeatother

\usepackage[
  unicode=true,
  pdfusetitle,
  bookmarks=true,
  bookmarksnumbered=false,
  bookmarksopen=false,
  breaklinks=false,
  pdfborder={0 0 0},
  pdfborderstyle={},
  backref=false,
  colorlinks=false
]{hyperref}

\hypersetup{
  colorlinks = true,
  urlcolor   = blue,
  linkcolor  = black,
  citecolor  = black
}

\makeatletter
\AtBeginDocument{%
  \@ifpackageloaded{hyperref}{%
    \providecommand\hyper@natlinkbreak[2]{#1}%
    \renewcommand\hyper@natlinkbreak[2]{#1}%
  }{}%
}
\makeatother

\usepackage[nameinlink,noabbrev]{cleveref}

\crefalias{subsection}{section}
\crefalias{subsubsection}{section}

\crefname{equation}{equation}{equations}
\Crefname{equation}{Equation}{Equations}

\crefname{figure}{figure}{figures}
\Crefname{figure}{Figure}{Figures}

\crefname{table}{table}{tables}
\Crefname{table}{Table}{Tables}

\crefname{section}{section}{sections}
\Crefname{section}{Section}{Sections}

\fancypagestyle{plain}{%
  \fancyhf{}%
}

\AtBeginDocument{%
  \pagenumbering{arabic}%
  \renewcommand{\thepage}{\arabic{page}}%
}

\newlength{\pagerightpull}
\newlength{\TBPageBottomInset}
\makeatletter
\AddToShipoutPictureFG{%
  \AtPageLowerLeft{%
    \ifTB@skipnextpagenum
      \global\TB@skipnextpagenumfalse
    \else
      \ifTB@inlandscape
        \put(\LenToUnit{\dimexpr\paperwidth-\TBPageBottomInset\relax},%
             \LenToUnit{\dimexpr\paperheight/2\relax}){%
          \makebox(0,0)[c]{\rotatebox{90}{\normalfont\normalsize\thepage}}%
        }%
        \ifTBInAppendix
          \put(\LenToUnit{\TBPageBottomInset},%
               \LenToUnit{\pagerightpull}){%
            \makebox(0,0)[bl]{\rotatebox{90}{\normalfont\normalsize\textcolor{gray}{Online Appendix}}}%
          }%
        \fi
      \else
        \put(\LenToUnit{\dimexpr\paperwidth/2\relax},%
             \LenToUnit{\TBPageBottomInset}){%
          \makebox(0,0)[c]{\normalfont\normalsize\thepage}%
        }%
      \fi
    \fi
  }%
}
\makeatother

\AtBeginEnvironment{appendices}{%
  \renewcommand{\thesection}{\Alph{section}}%
  \renewcommand{\thesubsection}{\thesection.\Alph{subsection}}%

  \titleformat{\section}{\Large\bfseries}{Appendix~\thesection.}{0.5em}{}

  \numberwithin{equation}{section}%

  \setcounter{figure}{0}%
  \counterwithout{figure}{section}%
  \renewcommand{\thefigure}{A.\arabic{figure}}%

  \setcounter{table}{0}%
  \counterwithout{table}{section}%

  \crefname{section}{appendix}{appendices}%
  \Crefname{section}{Appendix}{Appendices}%
  \crefalias{subsection}{section}%
  \crefalias{subsubsection}{section}%

  \pagestyle{fancy}
  \fancyhead[L]{\hspace*{\TBHeaderPad}\textcolor{gray}{Online Appendix}}

  \fancypagestyle{plain}{%
    \fancyhf{}%
    \fancyhead[L]{\hspace*{\TBHeaderPad}\textcolor{gray}{Online Appendix}}%
  }%
}

\usepackage{scalars}

 \title{Experimentally Validated Digital Twins: \\
Evaluating Workforce Training Programs in the Age of Big Data\thanks{Any opinions and conclusions expressed herein are those of the authors and do not represent the views of the U.S. Census Bureau. The Census Bureau has ensured appropriate access and use of confidential data and has reviewed the results for disclosure avoidance protection (Project 7519874: CBDRB-FY22-CES014-006, CBDRB-FY23-CES004-039, CBDRB-FY25-CES027-004, CBDRB-FY25-CES027-008, CBDRB-FY25-CES027-010, CBDRB-FY26-CES004-005, CBDRB-FY26-CES004-007, CBDRB-FY26-CES027-005, CBDRB-FY26-CES027-002, CBDRB-FY26-CES027-003, CBDRB-FY26-CES027-004, CBDRB-FY26-CES030-007, and CBDRB-FY27-CES027-004). James Reisinger worked at Opportunity Insights while contributing to this project. Raj Chetty leads Opportunity Capital, an organization that builds on the methods developed in this paper to invest in training programs. We are very grateful to Ruhi Baichwal, Corey Gelb-Bicknell, Juhui Jin, Josephine Kim, Samir Rein, and other Opportunity Insights pre-doctoral fellows for outstanding research assistance. We used AI to edit and proofread this manuscript. We thank Nathan Hendren, Guido Imbens, Yechan Park, Jonathan Roth, Jesse Shapiro, Davide Viviano, Joe Winkelmann, and Yiqing Xu for helpful comments and discussions. We are grateful to Richard Hendra and Kelsey Schaberg at MDRC and Mark Elliott and Anne Roder at Economic Mobility Corporation for providing data from randomized evaluations and answering questions about the programs and their implementation. We also thank Year Up United, Per Scholas, and Project QUEST for providing additional program records, and Social Finance for helping facilitate data access. This research was funded by the Andrew Carnegie Foundation, Chan Zuckerberg Initiative, Gates Foundation, Overdeck Family Foundation, Samvid Philanthropies, and the Walton Family Foundation.}}
\author{Raj Chetty, Harvard University and Opportunity Insights \\
Jamie Fogel, Harvard University and Opportunity Insights \\
Lawrence F. Katz, Harvard University and Opportunity Insights \\
Kadeem Noray, Harvard Business School\\
Sonya Porter, U.S. Census Bureau \\
James Reisinger, Meta}
\date{\vspace{0.5cm} October 2026}
 
\begin{document}
\onehalfspacing
\maketitle
\vspace{-0.5cm}
\thispagestyle{empty}

\begin{abstract}
\noindent Modern program evaluation prioritizes evidence from randomized controlled trials (RCTs) because prior research found that observational estimates of program impacts often diverge substantially from experimental benchmarks. We revisit this finding --- which was based on small survey datasets --- in the era of large-scale administrative data. Using anonymized income tax records covering all American adults between 1999 and 2025, we estimate the effects of workforce training programs on earnings by using machine learning models to predict counterfactuals from rich data on prior work histories, demographics, and residential locations. Rather than testing whether our observational estimates are unbiased, we use \NValidationBenchmarks{} RCTs to estimate the distribution of their bias. We find that our machine learning estimates closely match the RCT estimates, with root mean squared error (RMSE) comparable to that of the RCTs themselves. Our method reduces RMSE by \ValidationDMLVsLaLondeRMSEReductionPct{}\% relative to parametric estimators used in prior work because our larger samples and richer covariates permit more flexible, higher-dimensional models of the counterfactual. Applying our method to a broader set of sites and enrollment years beyond the experimental sample, we find that treatment effects on earnings vary substantially across locations and over time, even within a given program. Programs that place more participants into target-sector jobs generate the largest earnings gains, with impacts that persist even after participants leave their initial jobs. These findings provide guidance for improving job training programs and, more broadly, illustrate a scalable approach to program evaluation that uses observational data to construct experimentally validated ``digital twins.''

 \end{abstract}

\newpage
\clearpage
\setcounter{page}{1}
\doublespacing

\section{Introduction}\label{introduction}

The modern paradigm for program evaluation prioritizes evidence from randomized experiments. For example, the What Works Clearinghouse --- the U.S. Department of Education's repository of evidence used to guide funding decisions on educational interventions --- reserves its ``strong'' evidence rating for randomized controlled trials (RCTs) \citep{whatworks2026essatiers}. This paradigm emerged from a large body of research showing that observational estimates of a program's impacts --- obtained by comparing participants to non-participants who chose not to enroll --- do not reliably replicate estimates obtained from experiments where enrollment is randomly assigned (e.g., \citealp{ashenfelter1985longitudinal}, \citealp{lalonde1986evaluating}, \citealp{fraker1987adequacy}, \citealp{heckman1998characterizing}, \citealp{dehejia1999causal}, \citealp{glazerman2003nonexperimental}, \citealp{imbens2025comparing}). These findings catalyzed the ``credibility revolution,'' establishing experimental and quasi-experimental designs as the primary basis for program evaluation (\citealp{angrist2010credibility}).

Although well-designed randomized experiments provide high-quality evidence, they are expensive and challenging to implement. As a result, very few programs have been evaluated experimentally in practice. For instance, less than \JobTrainingRCTShareUpperBoundPct{} of the roughly \NUSJobTrainingPrograms{} job training programs currently operating in the United States \citep{gable2023workforce} have been evaluated in a randomized trial. Quasi-experimental sources of variation in program participation (such as sharp cut-offs in program eligibility) are also difficult to uncover and typically identify program impacts only for specific sub-groups of participants.

In this paper, we revisit the question of whether observational estimators can yield credible estimates of treatment effects, focusing on the canonical job training context studied in much of the prior methodological literature. Our re-analysis is motivated by the fact that prior work evaluating observational estimators was based on relatively small survey datasets with limited information. For instance, the main dataset used in the widely cited studies by \citet{lalonde1986evaluating} and \citet{dehejia1999causal,dehejia2002propensity} constructs a control group for an experimental job training program from a national sample of \NLaLondeComparisonSample{} adults from the cross-sectional Current Population Survey. Because it is difficult to find non-participants who match participants on all dimensions that matter for future earnings (e.g., age, location, prior earnings dynamics) in such datasets, researchers had to use relatively low-dimensional parametric models that extrapolate across individuals to construct counterfactuals.

Today, we have access to much larger administrative datasets that cover entire populations rather than small subsamples and have rich longitudinal data on many characteristics. Furthermore, machine learning methods make it feasible to construct flexible, high-dimensional models of counterfactual outcomes from such data. Can observational estimators provide a scalable, credible approach to program evaluation in the age of big data?

We address this question using an anonymized dataset constructed by linking two primary sources of data: (1) workforce training program records covering \NProgramParticipantsTotal{} participants and (2) income tax records covering the entire U.S. population, focusing on the \NPopulationAdults{} individuals who were 18 or older at some point between 1999 and 2025.

We begin our analysis by constructing new observational estimates of the impacts of workforce training programs on individuals' earnings. We use a machine learning model to predict each program participant's counterfactual earnings had they not participated in that training program. We then estimate treatment effects by comparing participants' realized outcomes to the predicted counterfactuals. Formally, we train gradient-boosted trees on the population of non-participants to predict future earnings based on a rich set of pre-program individual characteristics: work histories, industry indicators, demographics, and residential Census tract characteristics. We also estimate a propensity score model for program participation using the same vector of covariates. Finally, we combine the outcome and propensity score models using the double/debiased machine learning (DML) estimator of \citet{chernozhukov2018double}.\footnote{\citet{chernozhukov2018double} focus on estimation and inference using DML, taking identification as given. We focus on evaluating how well such methods identify treatment effects by comparing them to RCTs.}

We evaluate the performance of the DML estimator by comparing its estimates to estimates from \NValidationBenchmarks{} randomized trials of workforce training programs. Our approach differs conceptually from the literature initiated by \citet{lalonde1986evaluating}, which has used RCTs primarily to test whether unconfoundedness (selection on observables) holds \citep{rosenbaum1983central}, and, more broadly, from much of the modern program evaluation literature, which seeks research designs that eliminate selection bias. We instead start from the premise that unconfoundedness may fail: even with rich administrative data, factors such as motivation that affect both program participation and earnings remain unobserved, potentially generating selection bias in observational estimates. We use RCTs to estimate the distribution of bias in our DML estimator across programs, and then use that distribution to assess uncertainty about the DML estimates in new settings where no experiment is available, as in \citet{bernard2025trust}.

The randomized trials we analyze were conducted between \TrialFirstYear{} and \TrialLastYear{}. They span \NValidationProgramModels{} distinct job training and employment programs, ranging from sectoral programs such as Year Up United and Per Scholas, to welfare-to-work programs such as the Los Angeles Enhanced Job Club, to transitional job and wage subsidy programs such as Los Angeles Paid Work Experience.\footnote{Some programs (e.g., Year Up United and Per Scholas) were evaluated in multiple trials at different sites or time points.} For each experimental sample, we construct a DML estimate of the treatment effect on mean annual earnings 2--5 years after program enrollment by applying the algorithm described above to the RCT treatment group. We then compare these DML estimates to the RCT estimates. Both estimates are computed using the same group of treated individuals (program participants); they differ only in how the control comparison group is constructed: the RCT uses individuals randomly assigned to the control group in the trial, while the DML estimate uses the machine learning prediction of the counterfactual.

We find that the DML estimates match the RCT estimates closely, with a correlation of \ValidationCorrelation{} across the \NValidationBenchmarks{} trials. We estimate the distribution of bias in the DML estimates based on the difference between the DML and RCT estimates, adjusting for sampling error in the estimates. The noise-adjusted root mean squared error (RMSE) of the DML estimates (due purely to bias) is \ValidationNoiseAdjustedRMSEDollars{}, and the total RMSE (including the average degree of sampling error in the DML estimates) is \ValidationDMLTotalRMSEDollars{}. The RMSE of the RCT estimates (due to sampling error) is \ValidationRMSRCTSEDollars{}. Hence, the degree of error in the DML estimates is slightly smaller than that in the estimates from randomized trials currently used to guide decisions. Intuitively, the DML estimator makes up for greater error due to bias by using much larger samples than RCTs to estimate counterfactual earnings, which reduces sampling error.

Building on this comparison, we define an observational estimator as an experimentally validated ``digital twins'' estimator if it has RMSE comparable to or smaller than that of RCTs in a given domain (e.g., workforce training). We use the term ``digital twins'' for an estimator that satisfies this criterion because it yields counterfactuals --- statistical twins constructed from data on non-participants --- as accurate as those one would obtain from random assignment. By this standard, our DML estimator is an experimentally validated digital twins estimator for the class of training programs we study.

The digital twins estimates are precise enough to detect economically meaningful variation in treatment effects across programs. The RCT treatment effect estimates for the programs we study range from \ProgramEffectMinimumDollars{} to \ProgramEffectMaximumDollars{}, with a noise-adjusted standard deviation of \ProgramEffectSignalSDDollars{}. We show that if bias is normally distributed across programs, intervals of $\pm$1.96 times the RMSE around the digital twins estimates have at least 95\% coverage on average. This margin of error of 1.96 times the RMSE, \ValidationDMLMarginOfErrorDollars{}, is smaller than the standard deviation of treatment effects, implying that the digital twins estimates can detect typical (one-standard-deviation) deviations of a program's treatment effect from the mean across programs.

In contrast, observational estimators used in prior work exhibit much more bias. The parametric specifications of \citet{lalonde1986evaluating} and \citet{dehejia1999causal} yield noise-adjusted RMSEs of \$\ComparisonOLSSmallLaLondeRMSE{} and \$\ComparisonOLSSmallDehejiaWahbaRMSE{}, respectively, across the same \NValidationBenchmarks{} trials --- approximately \PriorSpecificationsRMSEMultiple{} times as large as the noise-adjusted RMSE of our DML estimator. Because their error due to bias alone exceeds the RMSE of the RCT estimates, these traditional specifications do not meet our ``digital twins'' criterion, consistent with the conclusions of prior work.

Simply applying the lower-dimensional specifications from prior work in larger samples does not yield a meaningful reduction in RMSE. Instead, the improvement comes from the combination of richer covariates --- especially longer work histories, which were unavailable in survey data --- and the capacity to estimate more flexible outcome models.\footnote{The propensity score model used for debiasing has little impact in our application; the outcome model alone yields RMSE comparable to that of the DML estimator.} Restricting our gradient-boosted trees to the covariates used by LaLonde yields much higher RMSE, as do lower-complexity models and gradient-boosted trees estimated in small samples comparable in size to RCT control groups (as in \citealp{bernard2025trust}). Together, these results show that big data improves observational estimates only when combined with models flexible enough to exploit its rich covariates and sample size.

Having validated our DML estimator in the experimental sample, we apply it to estimate the impacts of programs in settings without experimental evidence. To do so, we assume that the bias in the digital twins estimate for a new setting is drawn from the same marginal distribution as the biases across the \NValidationBenchmarks{} trials. Under this assumption, the \ValidationNoiseAdjustedRMSEDollars{} noise-adjusted RMSE estimated using the experimental validation sample can be used as a measure of uncertainty --- arising from potential selection bias rather than sampling error --- for the estimated effect in that setting.

Applying our method to new enrollment years and sites for programs that were previously evaluated in a randomized trial, we find that program impacts vary substantially over time and across sites. For example, a randomized trial of the Year Up United job training program conducted in \PACEAssignmentYears{} found that it increased participants' annual earnings by \YearUpPooledRCTEffectDollars{} (s.e. = \YearUpPooledRCTEffectSEDollars{}). Our digital twins approach yields a very similar treatment effect estimate of \YearUpPooledTwinsEffectDollars{}. However, applying the digital twins estimator to estimate earnings impacts for other enrollment years reveals that average impacts change substantially over time: they rose from \YearUpEarlyCohortEffectDollars{} for individuals who enrolled in Year Up United in \YearUpObsFirstCohort{}, when our records begin, to \YearUpPeakCohortEffectDollars{} for those who enrolled in \YearUpPeakCohortYear{} (shortly before the COVID-19 pandemic), before declining slightly to \YearUpLateCohortEffectLowerBoundDollars{} for those who enrolled in \YearUpLateCohortYear{}. Effects also vary significantly across locations: in some cities, Year Up United has large and rapidly growing treatment effects, while in others, effects are considerably smaller and show no systematic growth over time. These differences are highly statistically significant, based on hypothesis tests that account for error both due to sampling variation and potential bias in the DML estimates. We find similar patterns of increasing treatment effects on average over time for other programs, consistent with learning-by-doing as programs mature and refine their approaches. At the same time, past performance does not guarantee future returns: treatment effects vary sharply across sites as programs scale.

The variability of treatment effects across sites and over time increases the gains from using site- and enrollment-year-specific digital twins estimates relative to estimates from RCTs, which are typically conducted at a single point in time in a subset of sites. For example, consider estimating the treatment effect of Year Up United in a site whose true treatment effect is one standard deviation higher than the mean across sites. Using the pooled estimate from the \PACEAssignmentYears{} randomized trial for this target would yield an RMSE of \YearUpPooledRCTTargetRMSEDollars{}, \YearUpPooledRCTTargetRMSEMultiple{} times as large as the \ValidationDMLTotalRMSEDollars{} RMSE of the site-specific digital twins estimate. More generally, this comparison illustrates that experimentally validated digital twins estimates can provide \textit{more} accurate evidence for a particular site and year than typical randomized experiments that currently earn a ``strong'' evidence rating from the What Works Clearinghouse \citep{whatworks2026essatiers}.

Finally, we use our broader set of digital twins estimates to investigate the mechanisms that make workforce training programs effective. We correlate the digital twins treatment effect estimates with measures of program characteristics, labor market conditions, and participant characteristics. We find two systematic patterns among the programs that generate the largest gains. First, modern ``sectoral'' job training programs that not only teach new skills but also actively place participants into internships and target-sector jobs in high-wage sectors generate substantially larger earnings gains than programs that simply provide training or only provide temporary employment subsidies for low-wage jobs. This pattern holds even within programs: Year Up United has much larger effects in sites and enrollment years where it places a larger fraction of its trainees in target-sector jobs. 

Second, programs have highly persistent treatment effects over time: the programs that generate large gains in earnings two years after enrollment continue to do so 10--15 years after enrollment, while those that have little initial impact typically have little impact a decade later as well. These gains persist even as participants change jobs: \JobSwitcherSharePct{}\% of Year Up United participants switch jobs at some point in the five years after enrollment, and these switchers continue to earn \YearUpSwitcherEarningsGapDollars{} per year more than their digital twins counterfactuals. Hence, the mechanism through which sectoral programs produce earnings gains is not simply by getting people into jobs that pay more at a given point in time, but by getting them on a career ladder that yields permanent payoffs even as individuals switch positions. These findings suggest that the impacts of such programs may be durable even if market conditions change.

We conclude that observational estimators can provide reliable, near-real-time evidence on programs' causal effects and yield new insights into what makes interventions most effective in the age of big data. Nevertheless, we view observational methods as a complement to, not a substitute for, randomized experiments. Our common bias distribution assumption is credible only when the target program is similar in structure and population to those in our validation sample --- in our case, workforce training programs serving working-age adults with pre-program work histories. Applying the digital twins approach to programs in other domains requires experimental estimates in those domains to measure the error in observational estimates.

This paper builds on and contributes to two literatures. The first is the methodological literature initiated by \citet{lalonde1986evaluating} that uses RCT benchmarks to assess the credibility of observational estimators. The literature to date has focused primarily on testing whether a given observational estimator is unbiased (e.g., \citealp{dehejia1999causal}, \citealp{imbens2025comparing}, \citealp{young2026reevaluating}). We instead use many RCTs to characterize the \textit{distribution} of biases across programs, which allows us to make inferences about the likely magnitude of bias and expected loss (RMSE) of observational estimators in new settings. This approach lets decision makers use observational estimators even when those estimators are biased, provided that the bias is small relative to the differences in treatment effects they seek to detect. Our method is related to the literature on sensitivity analysis, which assesses robustness to violations of unconfoundedness under exogenously specified assumptions about the degree of selection on unobservables (e.g., \citealp{imbens2003sensitivity}, \citealp{altonji2005selection}, \citealp{oster2019unobservable}, \citealp{cinelli2020making}); we instead estimate the magnitude of such violations using experimental benchmarks.

Our analysis is most closely related to recent studies that have used multiple RCTs to estimate distributions of bias in other settings. \citet{bernard2025trust} study randomized trials across a range of domains in developing countries and find that observational estimators --- including DML --- exhibit substantial bias relative to the dispersion of treatment effects when implemented in the typically small datasets available to experimental researchers. \citet{gordon2023close} find that DML estimators exhibit substantial bias when compared to estimates from randomized trials of advertisements using large-scale data from Facebook, where ads are targeted using bidding information unavailable to the researchers. In contrast, our analysis demonstrates that flexible estimators trained in large samples on rich information directly relevant to the selection process (e.g., prior earnings histories) can provide estimates with reliability comparable to experiments. These differences suggest that observational estimators can be made reliable by building richer datasets, echoing the conclusion of \citet{heckman1997matching} and \citet{heckman1998characterizing}.\footnote{Consistent with this interpretation, \citet{gordon2023close} note that the bias in observational estimators in their advertising application is fundamentally a ``data problem'': if researchers had access to the detailed bidding data that determine how ads are targeted, selection on observables likely would be satisfied and the observational estimators would be unbiased.}

Second, we contribute to the growing literature evaluating workforce training and other programs to increase economic mobility. Several studies have reported experimental evidence on the impacts of individual training programs (e.g., \citealp{maguire2010tuning}, \citealp{fein2022benefits}, \citealp{katz2022sectoral}, \citealp{yusim2025effects}), while others have reported observational estimates using matching methods (e.g., \citealp{andersson2024federal}, \citealp{NBERw34174}). \citet{holzer2026workforce} and \citet{roodman2026evidence} present recent overviews and meta-analyses of evaluations of U.S. workforce programs; \citet{card2018works} provide a cross-country meta-analysis of evaluations of training and other active labor market programs. We contribute to this rich body of work by analyzing additional programs, providing longer-term estimates of earnings impacts (up to 15 years post-enrollment), documenting how treatment effects evolve as programs scale and adapt, and using the larger set of estimates to shed light on program features associated with effective workforce training. More broadly, our analysis shows that certain (but not all) sectoral workforce training programs generate larger earnings gains than most other educational interventions, suggesting that well-designed training programs are among the most effective tools for increasing upward mobility (see \citealp{chetty2026charterschools} for a comparison of impacts across interventions).

The paper is organized as follows. Section~\ref{data} describes our data sources. Section~\ref{estimand-and-estimation} presents our DML estimator. Section~\ref{experimental-validation} compares the DML estimates to RCT estimates, establishes that our estimator meets the digital twins criterion, and identifies the sources of its improvement over prior estimators. Section~\ref{applications} applies the estimator to estimate impacts in settings without experimental evidence. Section~\ref{program-features-associated-with-effective-training} uses these estimates to characterize the features of effective training programs. Section~\ref{conclusion} concludes by discussing implications for program evaluation more broadly. \section{Data}\label{data}

We use information from two data sources: workforce training program records and federal income tax records. Individuals are linked across these datasets using protected identification keys (PIK) generated by the Census Bureau for each dataset \citep{wagner2014person}. \ProgramPIKRatePct{}\% of individuals in the workforce training program datasets are successfully assigned a PIK and linked to tax records. All analyses are conducted within the Census Bureau's secure research environment using datasets stripped of identifiers, and all reported statistics are reviewed to protect confidentiality.  This section describes our data sources and defines the key variables we use in our analysis.

\subsection{Data Sources}\label{program-records-vcodex}

\textit{Workforce Training Program Data.} We obtain experimental records from randomized evaluations of workforce training programs and observational records for participants outside the experimental samples. The records identify the program, site (the location where the program was implemented), participants' enrollment dates, training track (e.g., IT or customer service), and other information collected by the provider or evaluator. The experimental records also include the date of random assignment, treatment status, and information on the corresponding RCT control group.

The experimental data are drawn from seven separate multi-site randomized evaluations conducted between \TrialFirstYear{} and \TrialLastYear{}, many of which evaluated multiple programs: WorkAdvance, the Sectoral Employment Impact Study (SEIS), Project QUEST, the Employment Retention and Advancement demonstration (ERA), the Subsidized and Transitional Employment Demonstration (STED), Year Up United PACE, and Year Up United Professional Training Corps (PTC).\footnote{Year Up United PACE and Year Up United PTC are separate randomized evaluations: PACE (Pathways for Advancing Careers and Education) evaluated Year Up United's stand-alone Core model, while PTC adapted that model for delivery in partnership with community colleges, using college instruction and facilities \citep{fein2018bridging,fein2022improvement}.} The original randomized evaluations were conducted by MDRC (WorkAdvance, ERA, and STED), Abt Associates (Year Up United PACE and PTC), Public/Private Ventures (SEIS), and Economic Mobility Corporation (Project QUEST). We obtained experimental records from MDRC for WorkAdvance, ERA, and STED; Economic Mobility Corporation for SEIS and Project QUEST; and Year Up United for Year Up United PACE and PTC.

Using these records, we construct a validation sample of \NValidationBenchmarks{} randomized trials that used mutually exclusive experimental samples. Each trial represents one site with experimental participants pooled across enrollment years.\footnote{We treat treatment arms that share a common control group as a single trial because our validation analysis effectively compares the observational control to the experimental control. We also combine some sites with very small samples into single trials. Specifically, we pool Year Up United PTC's three sites, Riverside Phase 2's Training Focused and Work Plus arms, and Los Angeles STED's Paid Work Experience and On-the-Job Training arms.} Together, these \NValidationBenchmarks{} trials span \NValidationProgramModels{} distinct job training and employment programs with differentiated selection processes.\footnote{There are \NValidationProgramModels{} distinct programs among our \NValidationBenchmarks{} benchmarks because four trials evaluated the same WorkAdvance model implemented by different providers, nine evaluated the Year Up United PACE program, and three evaluated a Texas ERA model with shared eligibility and recruitment rules \citep{tessler2014meeting,fein2018bridging,hendra2010effective}. Each of the remaining 21 trials evaluates a different program.} Although these trials are not a random sample of workforce training programs, they cover a wide range of interventions and selection processes that serve a broad range of program participants, from young adults entering the labor force to older displaced workers to individuals transitioning out of welfare. In total, there are \NValidationTreatedParticipantsTotal{} program participants in the randomized trials we analyze.

Observational records outside the experimental samples were obtained directly from Year Up United, Per Scholas, and Project QUEST, covering participation between \FirstYearUpObsYear{} and \LastYearUpObsYear{}. There are \NObservationalProgramParticipantsTotal{} program participants in the observational samples in total. %

\textit{Population Tax Data.} We use data from federal income tax records housed at the U.S. Census Bureau from 1999--2025, as in \citet{chetty2026opportunity}.\footnote{Here and below, a year refers to a tax year (i.e., the calendar year in which the income was earned).} The tax records include W-2 information returns filed by employers, individual income tax returns (Form 1040), and other information returns such as Forms 1099-DIV and 1099-INT. We also use employer and demographic characteristics from other data sources, which we discuss below.

\subsection{Variable Definitions}\label{tax-and-census-records-vcodex}

\textit{Earnings.} We define annual earnings as the sum of wage earnings reported on W-2 forms across all employers in a tax year. We code earnings as zero in years in which the administrative data are available but an individual receives no W-2. Hence, there are no individuals in the sample with missing earnings data by construction. Note that our baseline earnings measure excludes self-employment income, which is not reported on a W-2. We express earnings in \DollarYear{} dollars, adjusting for inflation using the Consumer Price Index (CPI-U), and top-code earnings at \$200,000.

\textit{Employers and industries.} The W-2 records identify the IRS employer identification number (EIN) for each employer from which an individual receives earnings, permitting us to measure the number of employers, earnings received from each employer, and changes in employers across years. We assign employers to industries by linking them to the corresponding year of the Longitudinal Business Database. Industries are measured using 2- and 4-digit NAICS codes associated with that firm in that year using the contemporaneous version of the NAICS code system (i.e., a firm's NAICS code may change over time as the NAICS system is updated).

\textit{Demographic characteristics.} We measure age and sex using the Numident. We measure marital status based on individual tax returns, defining an individual as married if they file as a married individual in a given year. We measure the number of children an individual has as the number of dependents they claim on Form 1040 in a given year.

\textit{Geographic characteristics.} We measure an individual's residential location at the Census tract level using addresses from tax and information returns, assigning each address to the corresponding tax year rather than the filing year. When location is unavailable, we use the most recent location observed in the prior five years. Tracts are defined using 2020 Census boundaries.

\textit{Other individual characteristics.} The linked records also contain information on college attendance, investment income, and other characteristics reported on individual income tax returns and third-party information returns. We define college attendance in a given year as having received a 1098-T form in that year. The 1098-T forms are available in calendar years 2011--2014 and 2016--2022. We define indicators for receiving interest or dividend income as indicators for receiving a 1099-INT or 1099-DIV form in a given year.

\textit{Year of training program enrollment.} We define the year in which an individual started a job training program based on the enrollment date in program rosters and the random-assignment date in experimental evaluation records. We refer to participants at a given program site who enroll in a given calendar year as an enrollment cohort.

 \section{Estimating Treatment Effects Using Observational Data}\label{estimand-and-estimation}\label{synthetic-digital-twins-and-estimation}

This section describes how we estimate treatment effects in observational data. We first define our target estimand, then describe our double/debiased machine learning (DML) estimator, and finally discuss how we implement that estimator in our data.

\subsection{Estimand}\label{experimental-estimand}

Our goal is to identify the average treatment effect of participating in a given job training program among those who choose to enroll, defined as the difference in their mean potential earnings with vs. without participation. Formally, let $D_{i}\in\{0,1\}$ represent an indicator for whether individual $i$ enrolls in the program of interest. Let $Y_i(1)$ denote $i$'s earnings (measured at some point after program enrollment) if they enroll in the training program and $Y_i(0)$ denote their earnings if they do not.\footnote{Throughout, we assume the standard no-interference condition that one individual's potential earnings do not depend on others' treatment status.} Our target estimand is the average treatment effect on the treated (ATT)
\begin{equation}\label{eq:estimand}
\begin{aligned}
\tau &= \mathbb{E}\!\left[Y_i(1)-Y_i(0)\mid D_{i}=1 \right] \\
     &= \mathbb{E}\!\left[Y_i(1)\mid D_{i}=1 \right] - \mathbb{E}\!\left[Y_i(0)\mid D_{i}=1 \right].
\end{aligned}
\end{equation}
Let $N_1=\sum_i D_i$ denote the number of treated individuals. The mean realized earnings of program participants, $\bar{Y}_{1}=\frac{1}{N_1}\sum_{i:D_i=1}Y_i$, provides an unbiased estimate of $\mathbb{E}\!\left[Y_i(1)\mid D_{i}=1\right]$. The central challenge in estimating $\tau$ is estimating $\mu_{0}=\mathbb{E}\!\left[Y_i(0)\mid D_{i}=1\right]$, the counterfactual mean earnings of program participants had they not participated. The rest of this section discusses how we estimate $\mu_{0}$ in observational data.

\subsection{Observational Estimator}\label{estimation-framework}

We estimate the counterfactual $\mu_{0}$ based on the expected earnings of an untreated individual with the same pre-treatment characteristics as each program participant. Let $X_i$ denote a vector of characteristics observed before treatment. The conditional mean of outcomes among individuals who do not participate in the training program is given by the outcome model
\begin{equation}\label{eq:outcome-regression}
m_0(x)=\mathbb{E}\!\left[Y_i\mid X_i=x,D_i=0\right].
\end{equation}

Using these outcomes among untreated individuals to measure the counterfactual $\mu_{0}$, we define the following observational counterpart to the ATT:
\begin{equation}\label{eq:tau-obs}
\tau^{\mathrm{OBS}}
=
\mathbb{E}\!\left[Y_i\mid D_i=1\right]
-
\mathbb{E}\!\left[m_0(X_i)\mid D_i=1\right].
\end{equation}

The observational comparison $\tau^{\mathrm{OBS}}$ equals the target ATT $\tau$ under two assumptions \citep{rubin1977assignment, rosenbaum1983central}: (1) selection on observables (unconfoundedness): conditional on $X_i$, untreated potential earnings are independent of program participation, $Y_i(0) \perp D_i \mid X_i$; and (2) overlap: the propensity to participate $\Pr(D_i = 1 \mid X_i) < 1$ over the support of $X_i$ among program participants.

In our setting, the overlap condition holds because we draw untreated comparison individuals from the full U.S. population, of which program participants are a very small fraction; estimated propensity scores are well below 1 throughout our estimation samples. Selection on observables, however, may not hold, because factors such as motivation that affect both program participation and earnings are unobserved even in our data. We therefore do not assume it holds; instead, we use randomized trials to measure the extent to which it fails in Section~\ref{experimental-validation}.

\textit{Machine Learning Estimator.} A natural plug-in estimator for $\tau^{\mathrm{OBS}}$ is the difference between realized and predicted earnings
\begin{equation}\label{eq:outcome-estimator}
\widehat{\tau}^{\mathrm{OBS}}
=\frac{1}{N_1}\sum_{i:D_i=1}\left\{Y_i-\widehat{m}_0(X_i)\right\},
\end{equation}
where $\widehat{m}_0(X_i)$ is an estimate of the outcome model among untreated individuals. When the covariate vector $X_i$ is high-dimensional, as in our setting, it is natural to use a flexible machine learning estimator for $\widehat{m}_0(X_i)$ that is regularized to prevent overfitting.\footnote{Although our comparison sample is very large, the dimensionality of $X_i$ rules out estimating $m_0(x)$ by exact matching on covariate profiles. Coarsening earnings in each of the five prior years into deciles and using only 4-digit industry in two prior years and county of residence yields roughly $10^{13}$ distinct covariate cells, about 100,000 times the number of working-age adults in the United States, even without including the remaining covariates. Flexible outcome models address this problem by pooling information across untreated individuals with similar, but not identical, characteristics.} Regularization minimizes prediction error through shrinkage but leads to biased estimates of $m_0(x)$, introducing bias in the plug-in estimate in~\eqref{eq:outcome-estimator}.\footnote{Intuitively, regularization pulls each individual estimate $\widehat{m}_0(X_i)$ toward a common central tendency, reflecting the fact that an extreme raw estimate is more likely to reflect sampling noise than a truly extreme conditional mean --- lowering prediction error on average but systematically compressing $\widehat{m}_0(x)$ toward the mean relative to the true $m_0(x)$.} Furthermore, this regularization bias converges to zero more slowly than the root-$n$ parametric rate because the number of parameters estimated in the model grows with sample size, unlike in traditional parametric models with a fixed functional form for the outcome model \citep{chernozhukov2018double}.

To address this bias, we augment the outcome model with a propensity score model of selection into program enrollment (\citealp{robins1994estimation}):
\begin{equation}\label{eq:propensity-score}
e(x)=\Pr\!\left(D_i=1\mid X_i=x\right).
\end{equation}
We estimate the propensity score $\widehat{e}(X_i)$ using a machine learning estimator with the same set of covariates as the outcome model.

Under overlap and standard regularity conditions on the estimation of $e$ and $m_0$, \citet{chernozhukov2018double} establish that the following ``double/debiased machine learning'' (DML) estimator is asymptotically unbiased for $\tau^{\mathrm{OBS}}$, root-$n$ consistent, and asymptotically normal:
\begin{equation}\label{eq:aipw}
\widehat{\tau}^{\mathrm{DML}}
=\frac{1}{N_1}\sum_{i:D_i=1}
\left\{Y_i-\widehat{m}_0(X_i)\right\}
-\frac{1}{N_1}\sum_{i:D_i=0}
\frac{\widehat{e}(X_i)}{1-\widehat{e}(X_i)}
\left\{Y_i-\widehat{m}_0(X_i)\right\}.
\end{equation}
The first term in~\eqref{eq:aipw} is the estimate based on the outcome model $\widehat{\tau}^{\mathrm{OBS}}$. The second term corrects the bias from regularization in the outcome model by measuring the average prediction error in the outcome model among untreated individuals, reweighting the untreated sample by the propensity odds $\widehat{e}(X_i)/[1-\widehat{e}(X_i)]$ so that its covariate distribution matches that of the treated sample. Intuitively, because regularization shrinks predictions toward the grand mean, the outcome model's prediction errors are generally not zero conditional on a given value of $X_i$. For example, if the outcome model underpredicts earnings for untreated individuals with characteristics $X_i$ like those of program participants (i.e., $Y_i-\widehat{m}_0(X_i)>0$), then $\widehat{\tau}^{\mathrm{OBS}}$ would overstate the treatment effect. The second term effectively estimates this degree of underprediction using a sample of untreated individuals with characteristics similar to program participants and subtracts it out.

We estimate both $m_0(x)$ and $e(x)$ using five-fold cross-fitting to avoid bias from overfitting \citep{chernozhukov2018double}. For each fold, we estimate the outcome and propensity score models on the other four folds to generate predictions for the observations in the held-out fold, so that every observation $i$ receives a predicted counterfactual outcome and propensity score from models estimated on a sample excluding $i$ itself.

\subsection{Empirical Implementation}\label{empirical-implementation}

We construct DML estimates of treatment effects $\widehat{\tau}^{\mathrm{DML}}_j$ at the program-site-enrollment-year level.\footnote{For cells in the experimental validation sample, we pool participants to match the treatment group in each randomized trial, combining enrollment years and, in some cases, sites or treatment arms, as discussed in Section~\ref{program-records-vcodex}.} We assign training program participants who enrolled in a given training program at the same site in the same year into the same cell, which we index by $j$. In each cell $j$, we construct an untreated comparison group based on all individuals in the population tax data who never participated at any site in the corresponding evaluation as follows.\footnote{When evaluating programs in which treatment was randomly assigned, we also exclude members of the RCT control group from the comparison group.} First, to align control individuals' outcomes with treated individuals' outcomes temporally, we randomly assign an enrollment year to those in the untreated comparison group uniformly from the set of enrollment years observed among participants in the corresponding evaluation. We then retain potential controls who (1) are alive in the assigned enrollment year, (2) reside in a county containing at least one participant in the evaluation, (3) fall between two years below the first percentile and two years above the 99th percentile of the participant age distribution in the evaluation, and (4) earn less than \$200{,}000 in each of the three preceding years.

We use actual enrollment years for participants and assigned enrollment years for untreated individuals to define covariates and outcomes in event time, with event year 0 corresponding to the actual or assigned calendar year of enrollment. We then estimate the outcome and propensity score models $m_0(x)$ and $e(x)$ separately for each cell.\footnote{For computational tractability, all estimates reported below use a random 10\% subsample of the eligible untreated comparison group, except the model-complexity analysis in Figure~\ref{fig:model_complexity}, which varies the subsample share. As Figure~\ref{fig:model_complexity} shows, estimates based on a 1\% subsample are very similar to those using the full population, so this subsampling has little effect on our estimates.}

We use three sets of covariates $X_i$ to estimate the outcome and propensity score models, all measured before program enrollment: (1) \textit{Work history}: total annual W-2 earnings and indicators for employment (positive W-2 earnings) in event years $-5$ through $-1$, mean earnings and the change in earnings over event years $-3$ through $-1$, and employer industry (both 2- and 4-digit NAICS codes) in event years $-2$ and $-1$; (2) \textit{Demographics}: sex, age in event year $0$, and marital status and presence of children in each of event years $-3$ through $-1$; and (3) \textit{Residential history}: state and county of residence in event year $-2$, together with characteristics of the residential Census tract in that year: poverty, unemployment, and high school graduation rates in 1990, 2000, and 2010; racial and ethnic composition in 2000; 3rd-grade reading and math achievement in 2015; and district-level standardized achievement measures from the Stanford Education Data Archive. Because estimation samples may include multiple enrollment years, we also include the calendar year corresponding to each individual's event year 0 as a covariate, thereby permitting the outcome and program participation models to vary with calendar time.

We estimate both the outcome and propensity score models using gradient-boosted decision trees (\citealp{friedman2001greedy}), implemented using LightGBM (\citealp{ke2017lightgbm}). Boosted trees repeatedly split the data at thresholds chosen by the model---for example, on prior earnings levels or industry---and combine many such trees, each trained on the residuals of the preceding trees, to minimize prediction error. We use gradient-boosted trees because prior work finds that they typically have the best predictive performance in tabular data among a wide range of machine learning models, partly because they handle skewed variables such as earnings well (\citealp{grinsztajn2022tree}; \citealp{shwartzziv2022tabular}; \citealp{mcelfresh2023neural}).\footnote{Consistent with these findings, when \citet{ahrens2025dml} compare twelve candidate machine learning models---including regularized regressions, random forests, and neural networks---in their labor market application, their model averaging procedure places nearly all weight on the boosted tree specifications.} Within the boosted tree family, we use LightGBM because it is optimized to handle training samples of hundreds of millions of observations and categorical variables such as detailed industry codes with a large number of values (\citealp{ke2017lightgbm}). We train the outcome model to minimize mean squared error loss and the propensity model to minimize binary log loss.

For each cell, we tune the hyperparameters that govern model complexity---such as the number of leaves per tree and the minimum number of observations per leaf---separately for the outcome and propensity score models. We select the hyperparameters that yield the best predictive performance in a held-out 20\% of the training data. Using these hyperparameters, we estimate both models using the cross-fitting procedure described in Section~\ref{estimation-framework} and finally construct the DML estimate in equation~\eqref{eq:aipw}. Following \citet{chernozhukov2018double}, we compute the standard error of each DML estimate as the square root of the empirical variance of its cross-fitted influence scores divided by the number of observations in the estimation sample.

Importantly, the DML estimates use no data from RCT control groups: the outcome model is trained entirely on non-experimental data for untreated individuals, and the propensity score model uses only pre-enrollment data on those individuals and program participants. Our observational models are therefore estimated and tuned without using information about the RCT estimates that we use to evaluate their performance in the next section.

 \section{Experimental Validation: Measuring the Error in DML Estimates}\label{experimental-validation}

In this section, we measure the degree of error in our DML estimates by comparing them to RCT estimates. We first compare the two sets of estimates descriptively and test whether the DML estimates are unbiased in every cell. We reject unbiasedness, but the more relevant question is whether the biases are small enough that the DML estimates remain useful for policy evaluation. We therefore define measures of error in observational estimates and a ``digital twins'' criterion for assessing whether those errors are small. We then apply these methods to estimate the degree of bias in the DML estimates and show that it is relatively small. Finally, we characterize the features of our estimator that account for its improved performance over the observational estimators used in prior work.

\subsection{DML vs. RCT Estimates}\label{validation-evidence}

We compare DML estimates with RCT estimates from a set $V$ of $J = \NValidationBenchmarks{}$ randomized trials spanning \NValidationProgramModels{} distinct program models, as discussed in Section~\ref{data}. Let $\overline{Y}_{1j}=\frac{1}{N_{1j}}\sum_{i\in j:D_i=1}Y_i$ denote the mean earnings of the $N_{1j}$ individuals randomly assigned to the treatment group (program participants) in randomized trial $j$, which we refer to as validation cell $j$.\footnote{The RCT treatment group consists of those \textit{offered} a position in the program (corresponding to an intent-to-treat effect) rather than those who enrolled. Because nearly all individuals assigned to treatment enrolled, we interpret the RCT treatment group means as measures of mean earnings among program participants. For example, \SEISTreatmentAttendancePct{} attended training in SEIS (\citealp[p.~10]{maguire2010tuning}), \WorkAdvanceProgramServicesReceiptPct{} received program services in WorkAdvance (\citealp[Table~2.3]{hendra2016encouraging}), \PACETreatmentEnrollmentPct{} enrolled in Year Up United PACE (\citealp{fein2018bridging}), \QUESTTreatmentSupportReceiptPct{} received support in Project QUEST (\citealp{roder2020nineyear}), and \PTCTreatmentEnrollmentPct{} enrolled in the Year Up United PTC evaluation (\citealp{fein2022improvement}).} Let $\overline{Y}^{\mathrm{RCT}}_{0j}=\frac{1}{N_{0j}}\sum_{i\in j:D_i=0}Y_i$ denote the mean earnings of the $N_{0j}$ individuals randomly assigned to the control group.

We estimate the RCT treatment effect $\widehat{\tau}_j^{\mathrm{RCT}}$ in cell $j$ as the difference in mean earnings between individuals in the treatment and control groups: $\widehat{\tau}_j^{\mathrm{RCT}}=\bar{Y}_{1j}-\bar{Y}^{\mathrm{RCT}}_{0j}$.\footnote{Some of the trials we analyze are stratified by pre-determined characteristics; we include strata fixed effects when estimating $\widehat{\tau}_j^{\mathrm{RCT}}$, but abstract from them here to simplify exposition.} We estimate the DML treatment effect $\widehat{\tau}^{\mathrm{DML}}_j$ in cell $j$ by applying the estimator in Section~\ref{estimand-and-estimation} to individuals assigned to the treatment group. Let $\widehat{\mu}^{\mathrm{DML}}_{0j} = \bar{Y}_{1j} - \widehat{\tau}_j^{\mathrm{DML}}$ denote the implied DML estimate of participants' mean counterfactual earnings had they not been treated. The DML counterfactual $\widehat{\mu}^{\mathrm{DML}}_{0j}$ plays the same role in the DML estimate that the control group mean $\bar{Y}^{\mathrm{RCT}}_{0j}$ plays in the RCT estimate.

To illustrate our approach, we begin by comparing DML and RCT estimates for three programs that span the range of populations served and treatments offered in our validation sample: Year Up United, which provides economically disadvantaged young adults with a \PACEProgramLength{} program pairing \PACEInstructionDuration{} of full-time classroom training with a six-month internship; Per Scholas, which provides a \SEISCourseWeeks{} information-technology course to unemployed and underemployed adults; and the Los Angeles Paid Work Experience program, which provides subsidized jobs to welfare recipients.

Panel A of Figure~\ref{fig:validation-examples} plots annual W-2 wage earnings for the treatment group ($\bar{Y}_{1j}$), RCT control group ($\bar{Y}^{\mathrm{RCT}}_{0j}$), and DML counterfactual ($\widehat{\mu}^{\mathrm{DML}}_{0j}$) in the \PACEAssignmentYears{} Year Up United PACE experiment.\footnote{PACE enrolled participants over two enrollment years at each site, with intake staggered across sites between January 2013 and August 2014. In the figure, we pool participants across the nine sites and two enrollment years, aligning them in event time (years relative to enrollment), and label event year 0 as 2014.}

Year Up United participants have low but steadily rising pre-enrollment earnings as they enter the labor market. Treated earnings fall during the enrollment year, as participants reduce work during their \PACEInstructionDuration{} of full-time training, and rise sharply thereafter relative to the RCT control group. The DML counterfactual closely matches the RCT control group. As a result, the DML estimator reproduces the RCT treatment effects. Panel B plots the RCT and DML treatment effect estimates, computed as the difference between the treatment group mean and the respective counterfactual estimates in Panel A. The DML estimate of the average treatment effect on earnings 2--5 years after program enrollment is \YearUpPooledTwinsEffectDollars{} (s.e. = \YearUpPooledTwinsEffectSEDollars{}), very similar to and statistically indistinguishable from the RCT treatment effect estimate of \YearUpPooledRCTEffectDollars{} (s.e. = \YearUpPooledRCTEffectSEDollars{}).

Panel C of Figure~\ref{fig:validation-examples} replicates Panel A for the Per Scholas randomized trial conducted in New York in 2005. Per Scholas participants enter the program with higher earnings (\PerScholasSEISEventYearMinusOneEarningsDollars{} on average in the year before treatment), but are on a downward wage trajectory. Earnings fall sharply in both experimental arms in the enrollment year, likely reflecting selection into the training program based on a transitory negative shock such as job loss --- an ``Ashenfelter dip'' widely documented in prior work on job training programs for adults (\citealp{ashenfelter1985longitudinal}). After enrollment, earnings rise much more sharply in the treated group than in the RCT control group and remain persistently higher for the next decade.

The DML counterfactual misses the transitory Ashenfelter dip in the year of program enrollment, but closely tracks the RCT control group after program enrollment. The estimated treatment effects are consequently once again very similar and statistically indistinguishable after program enrollment: an estimated treatment effect 2--5 years after program enrollment of \PerScholasSEISTwinsEffectDollars{} (s.e. = \PerScholasSEISTwinsEffectSEDollars{}) based on the DML estimator versus \PerScholasSEISRCTEffectDollars{} (s.e. = \PerScholasSEISRCTEffectSEDollars{}) from the experiment.

The Los Angeles Paid Work Experience program (Panel D), evaluated in an RCT in 2013, serves a substantially more disadvantaged population whose earnings fall steeply before entry, again reflecting selection into the program. This program has much smaller treatment effects on earnings. The estimated treatment effect 2--5 years after program enrollment is \STEDLAPWERCTEffectDollars{} (s.e. = \STEDLAPWERCTEffectSEDollars{}) based on the RCT. The DML counterfactual again misses part of the enrollment year dip, but tracks the control group closely in post-enrollment years, yielding a DML estimate of \STEDLAPWETwinsEffectDollars{} (s.e. = \STEDLAPWETwinsEffectSEDollars{}) on mean earnings 2--5 years after enrollment, similar to the RCT estimate.

Together, these examples indicate that the DML estimates exhibit relatively little bias across programs with sharp differences in earnings levels, pre-enrollment earnings trajectories, and treatment effects. The only systematic discrepancy arises in the enrollment year, where the DML counterfactual does not capture the sharp Ashenfelter dip observed in programs targeted at adults. This pattern is consistent with the fact that many adults enroll in job training programs because they experienced an unanticipated transitory shock such as a job loss. Because the shock is unanticipated, the pre-enrollment variables $X_i$ used to construct the DML counterfactual are unable to predict it, causing the model to overpredict earnings in the year of enrollment --- a violation of selection on observables. However, since such shocks are transitory, they contain little information about later earnings. As a result, the RCT control group's earnings revert to the path predicted by pre-enrollment covariates $X_i$, and the DML estimator matches the RCT estimates thereafter.

We generalize these case studies to examine bias across all \NValidationBenchmarks{} randomized trials. Figure~\ref{fig:validation-scatter} plots the RCT estimates against the DML estimates of treatment effects on mean earnings 2--5 years after program enrollment for the \NValidationBenchmarks{} trials.\footnote{We omit the first year after enrollment because many participants remain in training for part of that year, so the second year is the first fully post-training year for many programs; we average over the subsequent four years to reduce the influence of transitory fluctuations in earnings (and do not go beyond because some of the trials do not presently have more than five years of post-enrollment data).}  Vertical bars show 95\% confidence intervals for the RCT estimates; Appendix Figure~\ref{fig:validation-scatter-both-ci} reports confidence intervals for both estimators.

The estimates are clustered around the 45-degree line, with a raw unweighted correlation of \ValidationCorrelation{}. The mean unweighted difference between the DML and RCT estimates over these trials is \ValidationMeanGapDollars{}, a small difference relative to the broad dispersion in the RCT treatment effects, which range from \ProgramEffectMinimumDollars{} to \ProgramEffectMaximumDollars{}.

The similarity between the DML and RCT estimates suggests that the DML estimator captures much of the variation in program impacts across trials. But the DML estimates are not unbiased: a joint test of the hypothesis that the DML estimates are unbiased in all \NValidationBenchmarks{} cells --- accounting for sampling error in both the DML and RCT estimates --- rejects with $p<0.001$.\footnote{Because the DML and RCT estimates use the same treatment group, $\widehat{\tau}_j^{\mathrm{DML}} - \widehat{\tau}_j^{\mathrm{RCT}} = \overline{Y}_{0j}^{\mathrm{RCT}} - \widehat{\mu}_{0j}^{\mathrm{DML}}$ (see equation~\eqref{eq:counterfactual-gap}). This difference is centered around the bias $\mathbb{E}[\widehat{\tau}_j^{\mathrm{DML}}] - \tau_j$, with variance $\mathrm{Var}(\widehat{\mu}_{0j}^{\mathrm{DML}}) + \mathrm{Var}(\overline{Y}_{0j}^{\mathrm{RCT}})$. We test the joint hypothesis that the bias is zero in all cells using the Wald statistic $\sum_{j\in V} (\widehat{\tau}_j^{\mathrm{DML}} - \widehat{\tau}_j^{\mathrm{RCT}})^2 / [\mathrm{Var}(\widehat{\mu}_{0j}^{\mathrm{DML}}) + \mathrm{Var}(\overline{Y}_{0j}^{\mathrm{RCT}})]$, which rejects with $p = \ValidationJointWaldPValue{}$.} Selection on observables therefore does not hold exactly in our setting. However, the key policy question is not whether it holds, but whether the resulting bias is small enough that the DML estimates remain useful in practice. Answering this question requires a metric for the error in the DML estimates, which in turn depends on the decision problem the estimates are used to inform. We develop such a metric in the next subsection.

\subsection{Measuring Error in Observational Estimates}\label{validation-design}

Consider a decision maker deciding whether to invest in a program $j$ based on an observational estimate of its impact $\widehat{\tau}_j$.\footnote{\label{fn:no-pooling}In formulating the decision problem as a function only of the estimate for program $j$, we rule out estimators that use information from other programs to form the estimate for program $j$, either by shrinking estimates toward the mean across programs, as in empirical Bayes approaches, or by recentering them by the estimated mean bias, as in \citet{bernard2025trust}. Such non-separable estimators can reduce average loss, but make a program's estimated impact depend on data from other programs, crediting or penalizing it for their results --- an undesirable property in our policy applications.} Suppose the decision maker faces quadratic (squared-error) loss from mis-estimating the program's impact:
\begin{equation}\label{eq:dml-mse}
L_j=\mathbb{E}\left[(\widehat{\tau}_j - \tau_j)^2\right],
\end{equation}
where $\mathbb{E}$ denotes the expectation over the observational estimator's sampling distribution in cell $j$.

The decision maker's average loss from using the observational estimator to evaluate programs in the experimental validation sample is $L = \frac{1}{J}\sum_{j\in V} L_j$. Motivated by this decision problem, we focus on the square root of average loss, i.e., the root-mean-squared-error (RMSE) of the observational estimator, as our metric to evaluate its performance:
\begin{equation}\label{eq:observational-rmse}
RMSE(\hat{\tau})=\left[\frac{1}{J}\sum_{j\in V}\mathbb{E}\left[(\widehat{\tau}_j - \tau_j)^2\right]\right]^{1/2}.
\end{equation}
Determining whether a given value of $RMSE(\hat{\tau})$ is small or large requires a benchmark for comparison. One natural benchmark is the RMSE of the RCT estimates: if an observational estimator's RMSE is comparable to that of RCTs, it provides a good substitute for experimental evidence; if it is much larger, the decision maker might choose to delay her investment decision and run a randomized trial to gather more information (as in \citealp{epanomeritakis2026statistical}).

Because the RCT estimates are unbiased, the RMSE of the RCT estimates equals the square root of their average sampling variance, \[
RMSE(\hat{\tau}^\mathrm{RCT})=\sigma_{\mathrm{RCT}}
=
\left[
\frac{1}{J}\sum_{j\in V}\sigma_j^{2,\mathrm{RCT}}
\right]^{1/2},
\]
where $\sigma_j^{2,\mathrm{RCT}} = \mathrm{Var}(\widehat{\tau}_j^{\mathrm{RCT}})$ denotes the sampling variance of the RCT estimate in cell $j$ (which we estimate based on the squared standard errors of the RCT estimates).

We formalize this benchmark as follows:

\medskip
\noindent\textbf{Definition 1.} An observational estimator $\hat{\tau}$ is an \emph{experimentally validated digital twins estimator} if its RMSE is less than or comparable to that of RCT estimates:\footnote{We use approximate inequalities to convey that this criterion should be thought of as a guideline, not a strict rule, analogous to the common practice of using a 5\% significance threshold as a guideline for hypothesis testing.}
\[
RMSE(\hat{\tau})\lesssim\,\sigma_{\mathrm{RCT}}.
\]

The RMSE also provides a basis for inference. Under the assumption that the bias $B_j = \mathbb{E}\left[\widehat{\tau}_j\right] - \tau_j$ of the observational estimator is normally distributed across cells, along with regularity conditions stated in Appendix~\ref{app:rmse-coverage}, the intervals $\widehat{\tau}_j \pm 1.96\,RMSE(\widehat{\tau})$ have asymptotic coverage of at least 95\% on average across cells.\footnote{For the reasons discussed in footnote~\ref{fn:no-pooling}, we do not recenter our estimates by mean bias $\mathbb{E}[B_j]$ and therefore use only the second moment of the bias distribution, $\mathbb{E}[B_j^2]$, in constructing confidence intervals. The resulting intervals have nominal asymptotic coverage when mean bias is zero and are weakly conservative when mean bias is nonzero (see Appendix~\ref{app:rmse-coverage}).} This result provides a simple way to assess whether an observational estimator is precise enough to detect economically meaningful effects. For example, letting $\sigma_{\tau}$ denote the standard deviation of true treatment effects across the randomized trials in the validation sample, the estimator can detect typical (one-standard-deviation) deviations of a program's treatment effect from the mean across programs if its margin of error, $1.96\,RMSE(\widehat{\tau})$, is smaller than $\sigma_{\tau}$.

\subsection{An Estimator for the RMSE of DML Estimates} 

If the true treatment effects $\tau_j$ were observable, we could calculate the RMSE in equation~\eqref{eq:observational-rmse} and test whether our DML estimator satisfies the digital twins criterion directly. In this subsection, we derive a feasible estimator for the RMSE of our DML estimates that uses the RCT estimates $\widehat{\tau}_j^{\mathrm{RCT}}$ as noisy but unbiased estimates of $\tau_j$.

Let $\sigma_j^{2,\mathrm{DML}}=\mathbb{E}\left[\Big(\widehat{\tau}^{\mathrm{DML}}_j - \mathbb{E}\left[\widehat{\tau}^{\mathrm{DML}}_j\right]\Big)^2\right]$ denote the sampling variance of the DML estimator and $B_j = \mathbb{E}\left[\widehat{\tau}^{\mathrm{DML}}_j\right] - \tau_j$ denote its bias. We can decompose the expected squared error $L_j$ into variance and bias components:
\begin{equation}\label{eq:dml-mse-decomposition}
L_j=\sigma_j^{2,\mathrm{DML}}+B_j^2.
\end{equation}
We estimate the sampling variance $\sigma_j^{2,\mathrm{DML}}$ using the squared standard error of the DML estimate, constructed as described in Section~\ref{empirical-implementation}. We estimate the bias of the DML estimate in cell $j$ based on its deviation from the RCT estimate:
\begin{equation}\label{eq:estimated-bias}
\widehat{B}_j = \widehat{\tau}_j^{\mathrm{DML}} - \widehat{\tau}_j^{\mathrm{RCT}}.
\end{equation}
$\widehat{B}_j$, the horizontal deviation of each point from the 45-degree line in Figure~\ref{fig:validation-scatter}, is an unbiased (but noisy) estimate of $B_j$ ($ \mathbb{E}\left[\widehat{B}_j\right] = B_j$) because random assignment guarantees that the RCT estimate is an unbiased estimate of $\tau_j$ ($ \mathbb{E}\left[\widehat{\tau}^{\mathrm{RCT}}_j\right] = \tau_j$).

To estimate the bias component of the decomposition given by equation~\eqref{eq:dml-mse-decomposition}, we define the average squared bias in the validation sample as:
\begin{equation}\label{eq:rmsb-target}
\phi^2 = \frac{1}{J}\sum_{j\in V}\!B_j^{2}.
\end{equation}
Simply substituting $\widehat{B}_j $ into equation~\eqref{eq:rmsb-target} leads to an overestimate of $\phi^2$ because of noise in the estimates of bias. To derive an estimator that adjusts for this noise variance, note that because the DML and RCT estimates use an identical treatment group, the treatment group mean cancels out of~\eqref{eq:estimated-bias}. The estimated bias therefore equals the difference between the DML and RCT counterfactuals:
\begin{equation}\label{eq:counterfactual-gap}
\widehat{B}_j
=
\bar{Y}^{\mathrm{RCT}}_{0j}-\widehat{\mu}^{\mathrm{DML}}_{0j}.
\end{equation}
Let $\varepsilon^{\mathrm{RCT}}_{0j} \equiv \bar{Y}^{\mathrm{RCT}}_{0j}-\mu_{0j}$ and $\varepsilon^{\mathrm{DML}}_{0j} \equiv \widehat{\mu}^{\mathrm{DML}}_{0j}-\mathbb{E}[\widehat{\mu}^{\mathrm{DML}}_{0j}]$ denote the sampling errors of these two counterfactuals, where $\mu_{0j}$ is the treated group's expected earnings absent program participation in cell $j$. We can then write
\begin{equation}\label{eq:bias-error-decomposition}
\widehat{B}_j
=
B_j+\varepsilon^{\mathrm{RCT}}_{0j}-\varepsilon^{\mathrm{DML}}_{0j}.
\end{equation}
Note that $\mathbb{E}\!\left[\varepsilon^{\mathrm{RCT}}_{0j}\right]=0$ by random assignment and $\mathbb{E}\!\left[\varepsilon^{\mathrm{DML}}_{0j}\right]=0$ by construction. We estimate $\widehat{\mu}^{\mathrm{DML}}_{0j}$ and $\bar{Y}^{\mathrm{RCT}}_{0j}$ using distinct samples, so their sampling errors are uncorrelated. It follows that the expected mean squared estimated bias equals the mean squared bias of interest plus the average sampling variances of the two counterfactuals:
\begin{equation}\label{eq:squared-gap}
\frac{1}{J}\sum_{j\in V}\mathbb{E}\!\left[\widehat{B}_j^{2}\right]  =\phi^{2} +\frac{1}{J}\sum_{j\in V}\operatorname{Var}(\varepsilon_{0j}^{\mathrm{RCT}}) +\frac{1}{J}\sum_{j\in V}\operatorname{Var}(\varepsilon_{0j}^{\mathrm{DML}}).
\end{equation}
We estimate the noise variances as the squared standard errors of the RCT control group mean and DML counterfactual in each cell. We then obtain an unbiased estimator for the average loss due to bias, $\phi^2$, by averaging the squared bias estimates across the $J$ cells and subtracting the variance estimates:
\begin{equation}\label{eq:signal-rmse}
\widehat{\phi}^{\,2}
=
\frac{1}{J}\sum_{j\in V}\left(
\widehat{B}_{j}^{2}
-\widehat{\operatorname{Var}}\!\left(\bar{Y}^{\mathrm{RCT}}_{0j}\right)
-\widehat{\operatorname{Var}}\!\left(\widehat{\mu}^{\mathrm{DML}}_{0j}\right)
\right).
\end{equation}
Finally, we combine the estimated average squared bias with the average sampling variance across the $J$ validation cells to estimate the total RMSE of the estimator:
\begin{equation}\label{eq:dml-total-rmse}
    RMSE(\widehat{\tau}^{\mathrm{DML}})=\left[
            \hat{\phi}^{\,2}
            + \frac{1}{J}\sum_{j\in V}
              \widehat{\mathrm{Var}}
              (\widehat{\tau}^{\mathrm{DML}}_j)
          \right]^{1/2}.
\end{equation}
In the next subsection, we apply this framework to quantify the bias and RMSE of the DML estimates and evaluate whether the DML estimator meets the digital twins criterion.

\subsection{RMSE of DML Estimates}\label{quantifying-bias}

Table~\ref{tab:validation-summary} presents estimates of the variance components used to estimate the mean squared bias and total RMSE of the DML estimator in equations~\eqref{eq:signal-rmse} and~\eqref{eq:dml-total-rmse}. The first row of the table shows that the root mean squared difference between the DML and RCT estimates is $\left[\frac{1}{J}\sum_{j\in V}\widehat{B}_{j}^{2}\right]^{1/2}= \ValidationRawRMSEDollars{}$. This estimate overstates the root mean squared bias $\phi$ of the DML estimator because $\widehat{B}_j$ contains sampling error in the RCT and DML counterfactual estimates in addition to bias (equation~\eqref{eq:squared-gap}). The root mean sampling variances of the RCT and DML counterfactuals are \ValidationControlNoiseDollars{} and \PreliminaryValidationDMLCounterfactualNoiseDollars{}, respectively (rows 2 and 3). Subtracting these sampling variances as in equation~\eqref{eq:signal-rmse} yields a noise-adjusted estimate of the root mean squared bias, $\widehat{\phi}=$ \ValidationNoiseAdjustedRMSEDollars{} (row 4).\footnote{The mean unweighted difference between the DML and RCT estimates reported in Section~\ref{validation-evidence}, \ValidationMeanGapDollars{}, is an estimate of the mean bias of the DML estimator, $\mathbb{E}[B_j]$. $\widehat{\phi}$ is larger because it estimates $\left(\mathbb{E}[B_j^2]\right)^{1/2}$, the typical magnitude of bias in a given cell, rather than permitting positive and negative biases to offset.} $\widehat{\phi}$ can be interpreted as the RMSE of the DML estimator when sampling variance vanishes and only bias remains; it represents a lower bound on the RMSE of the DML estimator in finite samples.\footnote{If we exclude the ERA welfare-to-work and STED subsidized-employment evaluations, the estimated root mean squared bias is \ValidationNoiseAdjustedRMSEExcludingERASTEDDollars{}, and we no longer reject the hypothesis that the DML estimates are unbiased. Thus, our baseline estimate of the degree of bias in DML estimates is likely conservative for more typical workforce programs that have a formal training component.}

To estimate RMSE in finite samples, we must add back in the error due to sampling variance in the DML estimates. As a first step, row 5 of Table~\ref{tab:validation-summary} reports the root mean sampling variance of the DML treatment effect estimates when the DML estimator uses exactly the same group of treated individuals as the corresponding RCTs in the validation sample. Combining this sampling variance with $\widehat{\phi}$ as in equation~\eqref{eq:dml-total-rmse} yields a total RMSE of \UnscaledValidationDMLTotalRMSEDollars{} (row 6).

In practice, the number of program participants in observational settings is typically larger than in randomized trials because (a) there is no need to randomize a subset of participants into a control arm and (b) trials are typically implemented on a subset of participants given budget and logistical constraints. The sampling variance of the DML estimate is approximately equal to the sampling variance of the treatment group mean, which scales inversely with the number of treated individuals.\footnote{Because the DML counterfactual is estimated from the full population data, its sampling error is a negligible component of the total sampling error in the treatment effect estimates. Consistent with this logic, the standard errors of the DML estimates are nearly identical to the standard errors of the treatment group means across our validation cells.} Evaluating DML at the treated sample sizes of our RCT validation cells would therefore overstate its sampling variance in an observational evaluation of the same cells. To account for this, we scale the sampling variance of each DML estimate by the ratio of the mean treated sample size across RCT validation cells to the mean treated sample size across observational site-by-cohort cells for the same programs.\footnote{Specifically, we multiply each DML standard error by $\sqrt{\bar{N}_1^{\mathrm{RCT}} / \bar{N}_1^{\mathrm{Obs}}}$, where $\bar{N}_1^{\mathrm{RCT}}$ is the mean number of treated individuals across validation cells and $\bar{N}_1^{\mathrm{Obs}}$ is the mean number of treated individuals across observational site-by-cohort cells for Year Up United, Project QUEST, and Per Scholas --- the programs for which we have both experimental and observational data in our sample.} Combining this size-adjusted sampling variance with $\widehat{\phi}$ yields our preferred estimate of the total RMSE, $RMSE(\widehat{\tau}^{\mathrm{DML}})=$ \ValidationDMLTotalRMSEDollars{} (row 7 of Table~\ref{tab:validation-summary}).

The RMSE of the RCT treatment effect estimates is $\widehat{\sigma}_{\mathrm{RCT}}=\ValidationRMSRCTSEDollars{}$ (row 8 of Table~\ref{tab:validation-summary}), slightly larger than our estimate of $RMSE(\widehat{\tau}^{\mathrm{DML}}) =$ \ValidationDMLTotalRMSEDollars{}. The DML estimates have lower RMSE than the RCT estimates despite having bias because they have less sampling error, for two reasons: (1) the DML counterfactual $\widehat{\mu}^{\mathrm{DML}}_{0j}$ is estimated with negligible sampling error because it is based on very large samples drawn from the full population data and (2) the treatment groups available in observational settings are larger than those typically available in randomized trials.\footnote{A DML estimator with sufficiently large sample sizes relative to those in RCTs will have lower RMSE than the RCT, provided that its mean squared bias $\phi^2$ is less than the mean squared sampling error of the RCT estimates $\sigma_{\mathrm{RCT}}^2$, which is the case in our application. Specifically, the DML estimator outperforms the RCT when $\bar{N}_1^{\mathrm{Obs}} / \bar{N}_1^{\mathrm{RCT}} > \left(\frac{1}{J}\sum_{j\in V}\sigma_j^{2,\mathrm{DML}}\right) / (\sigma_{\mathrm{RCT}}^2 - \phi^2)$, which equals approximately \DMLRCTBreakEvenSampleSizeRatio{} in our setting --- similar to the ratio of observational vs.\ RCT treatment group sample sizes for Year Up United, Per Scholas, and Project QUEST. For programs operating at larger scale, the observational to RCT sample size ratio would presumably be larger, and the DML estimator would accordingly have lower RMSE.} Hence, by the standard of Definition~1, the DML estimator is an experimentally validated digital twins estimator for the class of training programs represented in our validation sample. We therefore refer to the DML estimates $\widehat{\tau}^{\mathrm{DML}}_j$ as ``digital twins'' estimates in the remainder of the paper.

Finally, we assess whether the DML estimates are precise enough to detect typical differences in treatment effects across programs. We estimate the standard deviation of true treatment effects across our \NValidationBenchmarks{} randomized trials, $\sigma_{\tau}$, by subtracting the average sampling variance of the RCT treatment effect estimates from the variance of their point estimates. The resulting noise-adjusted estimate is $\widehat{\sigma}_{\tau}=$ \ProgramEffectSignalSDDollars{} (row 9 of Table~\ref{tab:validation-summary}). The DML estimator's margin of error, $1.96 \times RMSE(\widehat{\tau}^{\mathrm{DML}}) =$ \ValidationDMLMarginOfErrorDollars{}, is smaller than $\widehat{\sigma}_{\tau}$, implying that the DML estimates can detect typical (one-standard-deviation) deviations of a program's treatment effect from the mean across programs.

\subsection{Sources of Improvement in RMSE Relative to Previous Observational Estimators}\label{performance-drivers}

Why does our observational DML estimator have low RMSE, whereas prior work since \citet{lalonde1986evaluating} has typically found that observational estimates deviate substantially from RCT benchmarks?

To determine why our conclusions differ, we begin by replicating LaLonde's specification in our context and calculating the RMSE of its estimates. For each of our \NValidationBenchmarks{} randomized trials, we regress post-enrollment earnings on an indicator for program participation and administrative data analogues of the covariates used by LaLonde in the Current Population Survey (CPS): W-2 earnings and employment status in the year prior to program enrollment, age and age squared, marital status, number of children, Black and Hispanic indicators (from the Decennial Census), and metropolitan area of residence.\footnote{The specification we estimate replicates that reported in column~10 of Table~5 of \citet{lalonde1986evaluating}, except that LaLonde controls for years of schooling and high-school-dropout status, which are unavailable in our data. To evaluate the importance of schooling controls, we additionally control for pre-program-enrollment college attendance using data from Form 1098-T, available in calendar years 2011--2014 and 2016--2022. Including this control has little impact on our estimates.} To approximate LaLonde's comparison sample construction, we randomly draw 136,000 individuals---the approximate size of the 1976 CPS---and apply eligibility restrictions in the spirit of his analysis: age below 55 and prior-year W-2 earnings below \$100,000. The resulting treatment effect estimates $\widehat{\tau}^{\mathrm{OLS}}$ have a noise-adjusted RMSE of \$\ComparisonOLSSmallLaLondeRMSE{} (Table~\ref{tab:estimator-comparison}, row 1, column 1), compared with \ValidationNoiseAdjustedRMSEDollars{} for our baseline DML estimator (row 3, column 4).\footnote{We focus on the noise-adjusted RMSE throughout this section to isolate the effect of different observational estimators on bias.} The OLS estimator controlling for variables available in the CPS thus does not meet the criterion for being a ``digital twins'' estimator: its noise-adjusted RMSE (a lower bound on total RMSE) exceeds the RMSE of the RCT estimates, $\widehat{\sigma}_{\mathrm{RCT}}=$ \ValidationRMSRCTSEDollars{}. This result is consistent with LaLonde's finding that such estimators do not reliably reproduce RCT estimates. Adding a second lag of earnings, following \citet{dehejia1999causal}, does not change this conclusion (row 2, column 1).

Our DML estimator $\widehat{\tau}^{\mathrm{DML}}$ differs from the OLS estimator $\widehat{\tau}^{\mathrm{OLS}}$ used by LaLonde and most other estimators in prior work \citep{imbens2025comparing} in four respects: it uses (1) a larger sample of control observations with greater overlap, (2) additional covariates $X_i$, (3) a more flexible, high-dimensional model of the counterfactual, and (4) an outcome and propensity score model together instead of using an outcome or propensity score model alone. We assess the contributions of these four factors to the improved performance of the DML estimator in turn.

\textit{Comparison Sample.} The comparison sample used by LaLonde and most prior studies to construct a control group differs from ours in two respects. First, it is much smaller: LaLonde had \NLaLondeComparisonSample{} observations, several orders of magnitude smaller than the \NPopulationAdults{} observations we use to construct our DML counterfactuals. Second, it has less overlap with the treated group: LaLonde started from a representative national sample with coarse age and wage restrictions (implemented in our replication as age below 55 and prior-year earnings below \$100,000), whereas we impose more targeted age and geographic restrictions to increase overlap with each treated sample (by further limiting the sample to individuals who live in the same counties as program participants and are of similar ages, as discussed in Section~\ref{empirical-implementation}).

To evaluate whether differences in comparison sample size and construction drive the reductions in RMSE, we re-estimate LaLonde's specification using our baseline sample (a 10\% random sample of the U.S. population with targeted overlap restrictions). This specification has an RMSE of \$\ComparisonOLSLargeLaLondeRMSE{} (Table~\ref{tab:estimator-comparison}, row 1, column 2) --- lower than the \$\ComparisonOLSSmallLaLondeRMSE{} RMSE in the original small sample, but substantially above the \ValidationNoiseAdjustedRMSEDollars{} of our baseline DML estimator. Most of the gain comes from the overlap restrictions rather than the change in sample size: the LaLonde estimator has an RMSE of \LaLondeNineMillionControlsRMSEDollars{} (not reported in the table) in a 10\% random sample of the full population with LaLonde-style age and earnings restrictions, similar to the RMSE in the small sample. These results show that sample construction matters, but does not by itself account for the improvement in the performance of our DML estimator.

\textit{Covariates.} Our DML estimator uses a much larger set of control variables than LaLonde was able to observe in the CPS. The LaLonde and Dehejia--Wahba datasets contained one or two years of prior earnings and employment information from surveys, demographic information similar to our data (age, marital status, and number of children), and coarse location information (metro area of residence). However, they did not contain information on earlier earnings trajectories, industry of employment, or the granular residential and local labor market characteristics we use here \citep{lalonde1986evaluating,dehejia1999causal,dehejia2002propensity}. Adding these characteristics to LaLonde's regression in a sample the same size as the one available to him reduces RMSE from \$\ComparisonOLSSmallLaLondeRMSE{} to \$\ComparisonOLSSmallExpandedRMSE{} (Table~\ref{tab:estimator-comparison}, rows 1 and 3, column 1)---still substantially above that of our baseline DML estimator. The benefits of richer covariates are greater when combined with DML and the population-scale sample. For example, going from LaLonde's covariate set to our richer set of covariates reduces RMSE from \$\ComparisonDMLLargeLaLondeRMSE{} to \ValidationNoiseAdjustedRMSEDollars{} in our population-scale sample (rows 1 and 3, column 4). The same change in covariates yields smaller gains when used in a linear model (rows 1 and 3, column 2). Hence, richer covariates are especially valuable when paired with flexible estimators and large samples.

To understand which covariates matter, we re-estimate the DML model in the population-scale sample using each of the seven combinations of the three covariate groups we use in our baseline model (defined in Section~\ref{empirical-implementation} above): work history, demographics, and residential history. Figure~\ref{fig:covariate_variation} reports the noise-adjusted RMSE of the DML estimate for each combination, computed as in equation~\eqref{eq:signal-rmse}. Controlling for demographics or residential history alone yields treatment effect estimates that exhibit very large bias, with RMSEs of \CovariateDemographicsOnlySignalRMSEDollars{} and \CovariateLocationOnlySignalRMSEDollars{}, respectively. Work history alone performs substantially better, with an RMSE of \CovariateWorkOnlySignalRMSEDollars{}. Adding either demographics or residential history to work history reduces the RMSE further, and the full set of covariates used in the baseline model yields the lowest RMSE, \ValidationNoiseAdjustedRMSEDollars{}. Work history is thus the most important predictor of counterfactual earnings, but demographics and residential history provide important complementary information.\footnote{Consistent with this conclusion, Appendix Figure~\ref{fig:shap-importance} shows that prior earnings variables contribute most to the fit of our baseline earnings outcome model, while demographic and geographic characteristics also rank highly. In this figure, covariate importance is measured using mean absolute SHAP values, which summarize the magnitude of each covariate's contribution to predicted counterfactual earnings.}

\textit{Model Flexibility.} The gains from using a more flexible machine learning model instead of a linear model can be seen by comparing columns 1--2 (OLS) of Table~\ref{tab:estimator-comparison} to columns 3--4 (DML). Applying DML to a sample with size and covariates comparable to those used by LaLonde yields an RMSE of \$\ComparisonDMLSmallLaLondeRMSE{} (row 1, column 3). This is lower than the RMSE of the linear model (row 1, column 1) but still substantially larger than the \ValidationNoiseAdjustedRMSEDollars{} RMSE obtained when applying DML in our population-scale sample with expanded covariates (row 3, column 4). Hence, model flexibility alone is not sufficient: DML achieves low RMSE only when combined with richer covariates and a larger sample. This finding echoes that of \citet{imbens2025comparing}, who show that DML does not yield substantial performance gains in LaLonde's original survey data. Conversely, richer covariates and a larger sample are not sufficient without a flexible model: OLS applied to our expanded covariates in the population-scale sample yields an RMSE of \$\ComparisonOLSLargeExpandedRMSE{} (row 3, column 2).

To characterize the gains from increasing model flexibility more continuously, we estimate gradient-boosted trees of varying complexity. Starting from our baseline specification, we reduce model complexity by scaling down the number of trees and the number of leaves per tree by a common factor; for example, a model of 33.3\% complexity uses one-third as many trees, each with one-third as many leaves, as our baseline specification. Figure~\ref{fig:model_complexity} reports the noise-adjusted RMSE for models of varying flexibility, estimated using samples of varying size: a 0.02\% random sample (comparable in size to the RCT control groups analyzed, e.g., by \citealp{bernard2025trust}); a 0.04\% random sample (comparable in size to the CPS sample analyzed by LaLonde); a 1\% sample; and the full population of \NPopulationAdults{} observations.\footnote{Our baseline estimator, which uses 100\% model complexity on a 10\% population sample, is not plotted in this figure; the full population sample (shown in teal) yields a similar RMSE at 100\% model complexity.}

The figure yields two results. First, reducing model complexity to 33.3\% of the baseline level has little effect on RMSE, but further reductions degrade performance sharply. For example, when using the full population data, RMSE rises from \ModelFullSampleFullComplexitySignalRMSEDollars{} at our baseline complexity level to \ModelFullSampleTenPctComplexitySignalRMSEDollars{} at 10\% complexity and \ModelFullSampleOnePctComplexitySignalRMSEDollars{} at 1\% complexity. Second, the gains from larger samples are greater when the model is more flexible. For example, going from a 0.02\% sample to the full population reduces RMSE by only \ModelPointZeroTwoToFullSampleThreePointThreePctComplexityRMSEReductionPct{}\% (from \ModelPointZeroTwoPctSampleThreePointThreePctComplexitySignalRMSEDollars{} to \ModelFullSampleThreePointThreePctComplexitySignalRMSEDollars{}) when using a model that is 3.33\% as complex as our baseline model. By contrast, the same increase in sample size reduces RMSE by \ModelPointZeroTwoToFullSampleFullComplexityRMSEReductionPct{}\% (from \ModelPointZeroTwoPctSampleFullComplexitySignalRMSEDollars{} to \ModelFullSampleFullComplexitySignalRMSEDollars{}) at our baseline level of model complexity.

\textit{Doubly Robust Estimation.} Finally, we ask whether the doubly robust structure of our estimator matters, comparing our DML estimates to estimates constructed from the outcome model alone and from propensity score reweighting alone. The estimator that uses only the outcome model has a noise-adjusted RMSE of \EstimatorOutcomeOnlySignalRMSEDollars{}, similar to the \ValidationNoiseAdjustedRMSEDollars{} of our baseline DML estimator. Debiasing the outcome model estimates using the propensity score correction does not change our estimates meaningfully. Intuitively, our outcome model has little regularization bias because our very large comparison sample allows us to form good predictions across the covariate space --- consistent with the fact that using the full population instead of a 1\% random sample has essentially no impact on the estimates (Figure~\ref{fig:model_complexity}). In contrast, propensity score reweighting alone performs substantially worse, with a noise-adjusted RMSE of \EstimatorPropensityOnlySignalRMSEDollars{}, perhaps because participation in any given program is rare and not well predicted by observables.

Together, these analyses show that our estimator's lower RMSE relative to earlier observational estimators comes primarily from using a higher-dimensional model of untreated earnings that can be estimated with the larger samples and richer set of covariates available today. These three features are complementary: in our setting, each is needed to achieve our baseline estimator's accuracy, but none is sufficient on its own. \section{Estimating Treatment Effects Beyond the Experimental Sample}\label{applications}

Having shown that our experimentally validated digital twins estimator performs well in replicating RCT estimates when they are available, we now apply it to settings without experimental evidence. We first state an assumption under which our estimates of the RMSE of the digital twins estimator in the validation sample provide information about the reliability of digital twins estimates in new settings. We then estimate treatment effects in additional sites and enrollment years. Finally, we discuss the implications of our findings for program evaluation, returning to the decision problem in Section~\ref{validation-design}.

\subsection{Common Distribution of Biases}\label{exchangeability-framework}

Our goal is to assess the accuracy of the digital twins estimator in a new cell $k\notin V$ for which no experimental evidence is available, drawing on information from cells in the experimental validation sample $V$, where both RCT and digital twins estimates are available. Formally, we seek to make inferences about the distribution of $B_k= \mathbb{E}\left[\widehat{\tau}^{\mathrm{DML}}_k\right] - \tau_k$ using the estimates of $\{B_j\}_{j\in V}$ from Section~\ref{quantifying-bias}.

We make the following assumption to link $B_k$ to the distribution of selection biases in the validation sample $\{B_j\}_{j\in V}$.

\medskip
\refstepcounter{assumption}\label{assumption:common-marginal-bias}
\noindent\textbf{Assumption~\ref{assumption:common-marginal-bias} (Common Distribution of Estimator Biases).} The target bias and the validation biases have a common marginal distribution $F_B$:
\[
B_j\sim F_B
\text{ for each }j\in V \cup \{k\}.
\]
Under Assumption~\ref{assumption:common-marginal-bias}, the distribution of biases in the experimental validation sample is informative about potential biases in the target cell $k$. In particular, the mean squared bias across the validation cells provides an unbiased estimate of the expected squared bias in the target cell:
\[
\mathbb{E}\!\left[B_k^2\right]
=
\mathbb{E}\!\left[
\frac{1}{J}\sum_{j\in V}B_j^2
\right].
\]
As a result, the noise-adjusted RMSE $\widehat{\phi}$, estimated in Section~\ref{quantifying-bias} as \ValidationNoiseAdjustedRMSEDollars{}, provides an estimate of $\sqrt{\mathbb{E}[B_k^2]}$, the root mean squared bias of a digital twins estimate in a new cell $k$. This estimate equals the RMSE of the digital twins estimate $\widehat{\tau}^{\mathrm{DML}}_k$ ignoring sampling error; we estimate the total RMSE of the estimate in cell $k$, $RMSE(\widehat{\tau}_k^{\mathrm{DML}})$, by adding back the cell-specific sampling variance:
\begin{equation}\label{eq:dml-cell-rmse}
RMSE\left(\widehat{\tau}_k^{\mathrm{DML}}\right)
=
\left[
\widehat{\phi}^{\,2}
+
\widehat{\sigma}_k^{2,\mathrm{DML}}
\right]^{1/2}.
\end{equation}
Note that Assumption~\ref{assumption:common-marginal-bias} restricts only the marginal distribution of estimator \textit{bias} $B_j$, not the distribution of \textit{treatment effects} $\tau_j$.\footnote{Assumption~\ref{assumption:common-marginal-bias} requires only that each bias have the same marginal distribution; it allows biases to be correlated across trials (e.g., across sites of the same program). Such correlation does not bias our estimate of $\mathbb{E}[B_k^2]$, the noise-adjusted mean squared bias $\widehat{\phi}^{\,2}$ from equation~\eqref{eq:signal-rmse}, but it reduces the effective number of independent validation cells and makes that estimate less precise. Inference about differences between cells requires additional assumptions about the joint distribution of their biases and sampling errors.} The programs we study are not a random sample of all training programs in the population and have widely varying programmatic approaches. We therefore remain agnostic about how treatment effects vary across cells. Our estimate of the treatment effect in cell $k$ ($\tau_k$) is based only on the outcomes of treated individuals in cell $k$ (and counterfactual predictions for those individuals constructed using the rest of the national data), consistent with the motivating decision problem in Section~\ref{validation-design}; see footnote~\ref{fn:no-pooling}.

From an economic perspective, Assumption~\ref{assumption:common-marginal-bias} requires that the bias arising from selection into program participation based on unobservable characteristics that affect earnings (e.g., motivation) has a common distribution across target and validation cells. This assumption is most plausible when selection into participation (conditional on observables $X_i$) is similar across cells. Selection is most likely to be similar in programs that deliver similar treatments to individuals with similar characteristics and use similar application processes. Motivated by these considerations, in our application, we focus on a set of programs that provide training for career preparation (rather than other objectives such as a broader liberal arts education), target working-age adults with prior labor market experience (rather than youth or others without work histories), and have open enrollment or only screen for basic competencies (rather than highly selective programs that screen applicants differently).

\subsection{Treatment Effect Estimates in Cells Without Experiments}\label{new-cohorts-sites}

To illustrate the use of our digital twins estimator in settings without experimental evidence, we estimate the effects of participation in Year Up United Core on earnings in additional enrollment years beyond those in the \PACEAssignmentYears{} PACE experiment.\footnote{We plan to report digital twins estimates for other programs beyond the experimental sample in subsequent versions of this manuscript.}

Panel A of Figure~\ref{fig:yearup-heterogeneity} plots the digital twins estimate $\widehat{\tau}^{\mathrm{DML}}_k$ of Year Up United's impact on earnings (measured two years after enrollment) by enrollment year, from \YearUpObsFirstCohort{} through \YearUpLateCohortYear{}.\footnote{We measure earnings in event year 2, two years after program enrollment, rather than the average earnings impact over event years \OutcomeStartEventYear{}--\OutcomeEndEventYear{} as in the validation analysis in Figure~\ref{fig:validation-scatter}, to maximize the set of enrollment years we can analyze. In practice, treatment effects are highly persistent: the unweighted correlation between treatment effects on earnings two years after program enrollment and treatment effects on mean earnings in years 3--10 after enrollment is \PersistenceCorrelation{} (s.e.\ = \PersistenceCorrelationSE{}) across cells for individuals who enrolled in programs before 2014.}  For people who enrolled in Year Up United in 2014, we estimate a treatment effect of \YearUpCohortEffectTwentyFourteenDollars{}, consistent with the RCT estimates reported in Section~\ref{validation-evidence}. However, this treatment effect changes substantially across enrollment years. Average treatment effects on earnings rose steadily from \YearUpEarlyCohortEffectDollars{} for those who enrolled in the program in \YearUpObsFirstCohort{} to \YearUpPeakCohortEffectDollars{} for those who enrolled in \YearUpPeakCohortYear{} --- perhaps as a result of ``learning by doing'' in the early years of the program. Because these differences are large relative to the estimated total RMSE of \ValidationDMLTotalRMSEDollars{}, they are unlikely to be explained by differential selection bias or noise: we reject the hypothesis that the treatment effects are the same in \YearUpObsFirstCohort{} and \YearUpPeakCohortYear{} with $p = \YearUpChangePValOne{}$, where the $p$ value accounts for both sampling error and potential bias in the estimates and is calculated as described in Appendix~\ref{app:rmse-coverage}. The earnings impacts then declined slightly during the COVID-19 pandemic to \YearUpLateCohortEffectLowerBoundDollars{} in \YearUpLateCohortYear{} ($p = \YearUpChangePValTwo{}$). 

Treatment effects vary substantially not just over time but across places. As an illustration, Panel B presents digital twins estimates of treatment effects on earnings by enrollment year in two cities where Year Up United operates, labeled City A and City B. The treatment effect in City A rises steadily from \YearUpCityAFirstCohortEffectDollars{} in \YearUpObsFirstCohort{} to \YearUpCityALateCohortEffectDollars{} in \YearUpLateCohortYear{} ($p=\YearUpChangePValThree{}$). In contrast, the treatment effects in City B do not change over the same period ($p=\YearUpChangePValFour{}$), averaging \YearUpCityBMeanEffectDollars{}.

\subsection{Implications for Program Evaluation}
\label{implications-program-evaluation}

The variation in treatment effects illustrated in Figure~\ref{fig:yearup-heterogeneity} for Year Up United---which we observe more generally for many other job training programs---increases the gains from using site- and enrollment-year-specific digital twins estimates rather than extrapolating from RCT estimates, which are typically obtained at a single point in time from a subset of sites due to budget and implementation constraints. For example, consider a decision maker seeking to estimate Year Up United's impact in a given year and site based on the average treatment effect $\widehat{\tau}^{\mathrm{RCT}}_{\mathrm{PACE}}$ from the \PACEAssignmentYears{} PACE randomized trial (e.g., as estimated in \citealp{fein2022benefits}). Let $\Delta_k=\tau_k-\mathbb{E}\!\left[\widehat{\tau}^{\mathrm{RCT}}_{\mathrm{PACE}}\right]$ denote the difference between the treatment effects in the target cell $k$ and the cells in the trial, which may arise from changes in treatment effects over time within a site or differences in treatment effects between sites (as illustrated in Figure~\ref{fig:yearup-heterogeneity}). The mean squared error of the RCT estimate is $\sigma^{2,\mathrm{RCT}}_{\mathrm{PACE}}+\Delta_k^2$, while the mean squared error of the digital twins estimate is $\phi^{2}+\sigma_k^{2,\mathrm{DML}}$, as shown in equation~\eqref{eq:dml-cell-rmse}, under Assumption~\ref{assumption:common-marginal-bias}.\footnote{This comparison differs from the one in Section~\ref{quantifying-bias}, where both the RCT and DML estimators were applied to the same cell. Here, we compare the MSE of the cell-specific DML estimates to that of the historical RCT estimate pooled across sites, extrapolated to a different site and enrollment year.} The digital twins estimate therefore has lower expected squared error when
\begin{equation}\label{eq:historical-other-site-breakeven}
\Delta_k^2
>
\phi^{2}
+\sigma_k^{2,\mathrm{DML}}
-\sigma^{2,\mathrm{RCT}}_{\mathrm{PACE}}.
\end{equation}

For Year Up United, this condition holds when the target site's true treatment effect differs from the pooled treatment effect in the \PACEAssignmentYears{} PACE experiment by more than \YearUpNewSiteBreakevenDollars{}.\footnote{The threshold is positive because the pooled PACE estimate is more precise than a typical single-trial estimate: by pooling \NYearUpPACESitesWords{} sites, it has a standard error of \YearUpPooledRCTEffectSEDollars{}, compared with \ValidationRMSRCTSEDollars{} for a typical trial in our validation sample. Absent extrapolation bias, the pooled PACE estimate would therefore be more accurate than a digital twins estimate for a single cell. However, this gain in precision by pooling sites is exactly what generates the cost of not capturing the heterogeneity in treatment effects across cells.} This threshold is small relative to the variation documented above: the noise-adjusted standard deviation of treatment effects across the \NYearUpPACESitesWords{} PACE sites during the \PACEAssignmentYears{} enrollment years alone was \YearUpPACESiteSignalSDDollars{}, and Figure~\ref{fig:yearup-heterogeneity} indicates that treatment effects within sites changed by much more than \YearUpNewSiteBreakevenDollars{} over time.

These calculations imply that site- and enrollment-year-specific digital twins estimates of Year Up United's impacts will typically have lower RMSE than the pooled historical RCT estimate. For example, consider estimating Year Up United's effect on mean annual earnings in a site where the true treatment effect is one standard deviation (\YearUpPACESiteSignalSDDollars{}) higher than the mean effect in the PACE experiment. The RMSE of the pooled PACE experimental estimate would be \YearUpPooledRCTTargetRMSEDollars{} for this target, \YearUpPooledRCTTargetRMSEMultiple{} times as large as the RMSE of the cell-specific digital twins estimate, which is \ValidationDMLTotalRMSEDollars{} for a cell of average size in the observational sample.

More generally, these calculations suggest that experimentally validated digital twins estimators could provide a more accurate basis for program evaluation than evidence from typical randomized trials (e.g., \citealp{whatworks2026essatiers}) given the variability of treatment effects over time and sites.\footnote{Of course, the digital twins estimates will not outperform all experimental estimates: a sufficiently large-scale RCT conducted in the site and year of interest would have lower RMSE than the digital twins estimator. Our point is that the digital twins estimates can outperform typically available estimates from RCTs.} \section{Mechanisms Underlying Effective Training Programs}\label{program-features-associated-with-effective-training}

In this section, we use our digital twins estimates to characterize the features of effective workforce training programs by correlating our digital twins treatment effect estimates with measures of program characteristics, labor market conditions, and participant characteristics. We first state an assumption under which the association between our DML estimates and characteristics identifies the associations of interest and then present empirical estimates.

\subsection{Identifying Associations Between Treatment Effects and Characteristics}\label{mechanisms-framework}

Our goal is to estimate the association between cell-level treatment effects $\tau_j$ and an observable characteristic $Z_j$, captured by the regression coefficient
\begin{equation}\label{eq:beta-z}
    \beta_{Z} = \frac{\mathrm{Cov}(\tau_j, Z_j)}{\mathrm{Var}(Z_j)}.
\end{equation}
To estimate $\beta_Z$ by using our digital twins treatment effect estimates $\widehat{\tau}^{\mathrm{DML}}_j$, we make the following assumption:

\medskip
\refstepcounter{assumption}\label{assumption:orthogonal-bias}
\noindent\textbf{Assumption~\ref{assumption:orthogonal-bias} (Bias Orthogonal to Characteristics).} The bias $B_j$ of the DML estimate in cell $j$ is orthogonal to the characteristic of interest $Z_j$:
\[
\mathrm{Cov}(B_j, Z_j) = 0.
\]
Under this assumption, $\mathrm{Cov}(\widehat{\tau}_j^{\mathrm{DML}}, Z_j) = \mathrm{Cov}(\tau_j, Z_j)$, since sampling error in $\widehat{\tau}_j^{\mathrm{DML}}$ is uncorrelated with $Z_j$. Hence, regressing $\widehat{\tau}^{\mathrm{DML}}_j$ on $Z_j$ identifies $\beta_Z$. Intuitively, if the digital twins estimator's bias systematically covaries with cell-level characteristics, then the association between our treatment effect estimates and characteristics could reflect that bias rather than differences in program impacts; Assumption~\ref{assumption:orthogonal-bias} rules out this possibility. We first report estimates assuming Assumption~\ref{assumption:orthogonal-bias} holds, and then present a bounding exercise showing that violations of Assumption~\ref{assumption:orthogonal-bias} are unlikely to be driving our conclusions.

\subsection{Empirical Estimates}\label{correlation-estimates}

We illustrate our approach by focusing on variation in treatment effects across sites and enrollment years in the Year Up United program.\footnote{We plan to characterize mechanisms using observational estimates for a broader range of programs in subsequent versions of this manuscript.} The digital twins estimator yields \NYearUpPlacementCells{} treatment effect estimates for Year Up United's Core and PTC programs, covering \NYearUpPlacementSites{} sites and enrollment years from \YearUpPlacementFirstCohort{} through \YearUpPlacementLastCohort{}. When we correlate these estimates with a broad range of program characteristics---such as the number of participants, sectoral focus, and local labor market conditions---one variable emerges as the strongest predictor of the heterogeneity in these effects: the share of program participants placed into full-time jobs in the target sector within four months of program completion.\footnote{Individuals who do not complete the program are defined as not being placed.}

Figure~\ref{fig:yearup-placement} illustrates this result by presenting a binned scatter plot of estimated earnings impacts against the share of participants placed into full-time target-sector jobs in each cell. Target-sector placement rates are strongly positively correlated with estimated impacts ($\text{correlation}= \YearUpPlacementCorrelation{}$). The average treatment effect is \YearUpLowPlacementEffectDollars{} per year among site-by-enrollment-year cells with placement rates below \YearUpLowPlacementThresholdPct{}\%, compared with \YearUpHighPlacementEffectDollars{} among cells with placement rates of \YearUpHighPlacementThresholdPct{}\% or higher. We estimate $\hat{\beta}_{Z} = \BetaZ{}$ per percentage point (s.e. = \$\BetaZse{}), implying that a one standard deviation increase in placement rates --- about \OneSDPlacement{} percentage points --- is associated with a \OneSDPlacementAssociatedEarningsImpact{} larger annual earnings impact.

The validity of our $\hat{\beta}_{Z} = \BetaZ$ estimate rests on Assumption~\ref{assumption:orthogonal-bias}; if cells with higher target-sector placement rates also have systematically more positive DML bias, then this estimate would overstate the association between treatment effects and placement rates. We assess whether violations of Assumption~\ref{assumption:orthogonal-bias} could produce the association observed in Figure~\ref{fig:yearup-placement} using a bounding exercise. Observe that $\hat{\beta}_Z = \beta_Z + \mathrm{Cov}(B_j,Z_j)/\mathrm{Var}(Z_j)$, and $|\mathrm{Cov}(B_j,Z_j)| \le \mathrm{SD}(B_j)\,\mathrm{SD}(Z_j)$. Hence the difference in estimated impacts associated with a one standard deviation increase in placement rates $Z_j$ can be overstated by at most $\mathrm{SD}(B_j)$. Under Assumption~\ref{assumption:common-marginal-bias}, $\mathrm{SD}(B_j)$ is at most $\widehat{\phi} = \ValidationNoiseAdjustedRMSEDollars{}$, the noise-adjusted RMSE estimated in Section~\ref{quantifying-bias}.\footnote{$\widehat{\phi}^{\,2}$ is an upper bound on $\mathrm{Var}(B_j)$ because it estimates $\mathbb{E}[B_j^2] = \mathrm{Var}(B_j) + \mathbb{E}[B_j]^2$.} Even in the extreme case in which bias is perfectly correlated with placement rates across cells, bias cannot fully account for the association we observe empirically: at most \ValidationNoiseAdjustedRMSEDollars{} (\ValidationNoiseAdjustedRMSEPercentageofCorrelation{}) of the \OneSDPlacementAssociatedEarningsImpact{} association between earnings impacts and placement rates can be attributed to bias in the digital twins estimates.

Since Year Up United's skills training curriculum is largely standardized across sites and enrollment years, the association between earnings impacts and placement rates suggests that the program's focus on placing participants into high-wage sectoral jobs underlies its success, rather than the training component alone. This conclusion is consistent with prior evidence that modern ``sectoral'' job training programs, which combine training with active placement into target-sector jobs, generate substantially larger earnings gains than traditional programs that provide training alone (\citealp{katz2022sectoral}; \citealp{holzer2026workforce}; \citealp{roodman2026evidence}). Our analysis shows that the same pattern holds even \textit{within} a given program, strengthening the conclusion that active job placement is a critical complement to training itself.\footnote{The association in Figure~\ref{fig:yearup-placement} does not establish a causal effect of target-sector placement rates on earnings impacts, because cells with higher placement rates could differ in other respects (e.g., participant selection, local labor demand, or implementation quality). A useful direction for future work would be to combine the digital twins estimates reported here with an exogenous source of variation in job placement rates---e.g., idiosyncratic shocks to relationships with local employers---to identify causal effects. Regardless of the underlying causal mechanism, target-sector job placement rates are a useful real-time predictor of subsequent earnings impacts.}

Although initial placement into a target-sector job is highly predictive of earnings gains, Year Up United's longer-term earnings gains --- which persist for at least 10 years after program enrollment, as observed in Panel B of Figure~\ref{fig:validation-examples} --- are not driven simply by placement into one specific high-paying firm. Only \JobStayerSharePct{}\% of Year Up United Core participants remain at the firm where they were initially placed five years after program enrollment. Figure~\ref{fig:job-industry-switchers} plots earnings trajectories for the \JobSwitcherSharePct{}\% of participants who switch jobs within five years after program enrollment. These individuals continue to earn \YearUpSwitcherEarningsGapDollars{} per year more than their digital twins counterfactuals five years after program enrollment, even after switching jobs. These findings indicate that Year Up United's persistent earnings gains reflect durable improvements in human capital and labor market opportunities that outlast any one job.

We conclude that effective job training programs increase earnings by placing participants onto a career ladder that yields higher earnings even as they change jobs. This mechanism suggests that the impacts of sectoral programs may be durable even as technology and economic conditions change. \section{Conclusion}\label{conclusion}

This paper has studied whether observational estimators can produce credible estimates of program impacts in the age of big data, focusing on the case of workforce training programs. We find that machine learning estimates constructed using tax data covering all Americans have root mean squared error that is comparable to that of existing RCT estimates and one-third as large as the standard deviation of true treatment effects across our \NValidationBenchmarks{} randomized trials. We term observational estimators with RMSE comparable to or smaller than that of RCTs experimentally validated ``digital twins'' estimators, because they identify non-participants whose predicted outcomes provide counterfactuals as good as those from random assignment. Such estimators were not feasible in prior program evaluations using observational data (e.g., \citealp{lalonde1986evaluating}), whose smaller survey data admitted only low-dimensional parametric models of the counterfactual. Large-scale administrative data with rich covariates now support much more flexible models that reduce selection bias substantially.

Applying our digital twins estimator beyond the experimental sample, we find that treatment effects vary sharply across sites and enrollment years even within a given workforce training program. Sectoral job training programs that successfully place participants into target-sector internships and jobs generate the largest earnings gains. These gains persist even after participants move out of their initial jobs, showing that effective training programs raise earnings by placing participants on higher-return career trajectories rather than simply placing them in a specific high-paying job.

Our findings suggest several directions for future work, both in applications to workforce training and in developing credible observational methods. In the context of workforce training, our digital twins estimator can be applied to any program with linkable participant records, making near-real-time impact evaluation feasible at scale for the \NUSJobTrainingPrograms{} programs currently operating in the United States. Applying the estimator developed here to measure the returns of these programs would allow policymakers to continually reallocate resources toward programs that generate the largest earnings gains, adjusting as their impacts change. Constructing estimates for a broader set of programs would also facilitate further research on the mechanisms that make workforce training most effective.

Methodologically, it would be useful to test whether going beyond the machine learning models considered here can further improve the performance of observational estimators. For example, the gradient-boosted trees we use exploit tabular administrative data, but do not incorporate any unstructured information, such as text descriptions of program participants' prior workplaces or neighborhoods. Incorporating such unstructured information using modern large language models could sharpen counterfactual predictions. It would also be useful to evaluate the performance of observational estimators in settings with more limited pre-program data. For example, the feasibility of developing digital twins estimators for subgroups that lack the rich work histories on which our predictions rely --- such as youth, immigrants, and other first-time labor market entrants --- is an important open question.

More broadly, our results point to a different perspective on the role of experimental evidence in the age of big data. Given the substantial heterogeneity in treatment effects across sites, subgroups, and time periods, a single RCT or quasi-experimental estimate at one point in time may be insufficient to guide policy decisions in other contexts. Randomized trials and quasi-experiments may be most valuable as tools to calibrate and validate observational estimators --- which can then generate the site-, enrollment-year-, and subgroup-specific estimates relevant to policy decisions --- rather than as the exclusive source of treatment effect estimates themselves. Using rich data to build observational estimators that meet the ``digital twins'' standard in other domains --- e.g., health, education, and criminal justice --- is a promising direction for future work. 
\newpage 

\providecommand{\BibliographyPath}{references}
%%%%%%%%%%%%%%%%%%%%%%%%%%%%%%%%%%%%%%%%%%%%%%%%%%%%%%%%%%%%%%%%%%%
%  This bbl file is created by aea.bst ver.3.1.1
%  The latest aea.bst is available at
%  <https://github.com/ShiroTakeda/econ-bst>

\newpage

\setcounter{figure}{0}

\begin{figure}[p]
\newcommand{\validationexamplesubcaption}[1]{%
  \parbox[t][2\baselineskip][t]{\linewidth}{\caption{#1}}%
}

\centering
\captionsetup{labelsep=newline}
\captionsetup[subfigure]{font={small,singlespacing},justification=centering}
\caption{\hspace{0.5em}Randomized Trial vs. Machine Learning Estimates: Event Study Comparisons}
\label{fig:validation-examples}
\begin{minipage}{\linewidth}
\centering
\begin{subfigure}[t]{0.49\linewidth}
\centering
\validationexamplesubcaption{Year Up United: Earnings Levels}
\includegraphics[width=\linewidth]{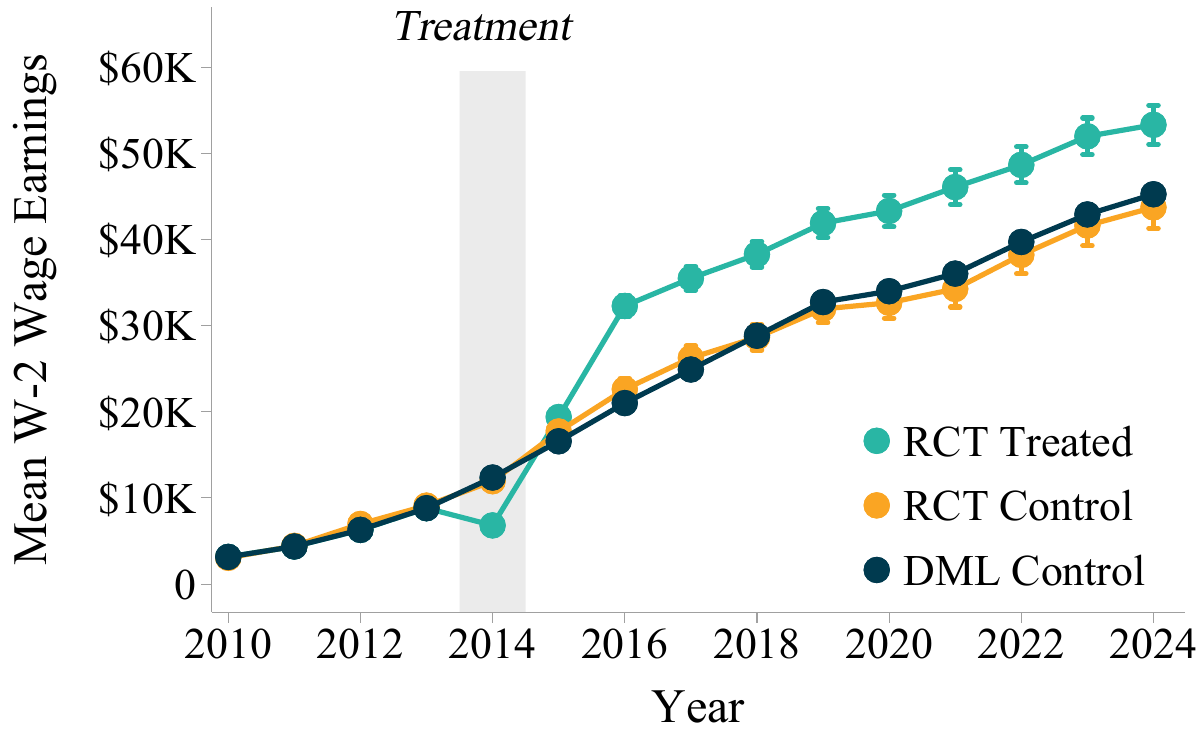}
\end{subfigure}
\hfill
\begin{subfigure}[t]{0.49\linewidth}
\centering
\validationexamplesubcaption{Year Up United: Treatment Effects}
\includegraphics[width=\linewidth]{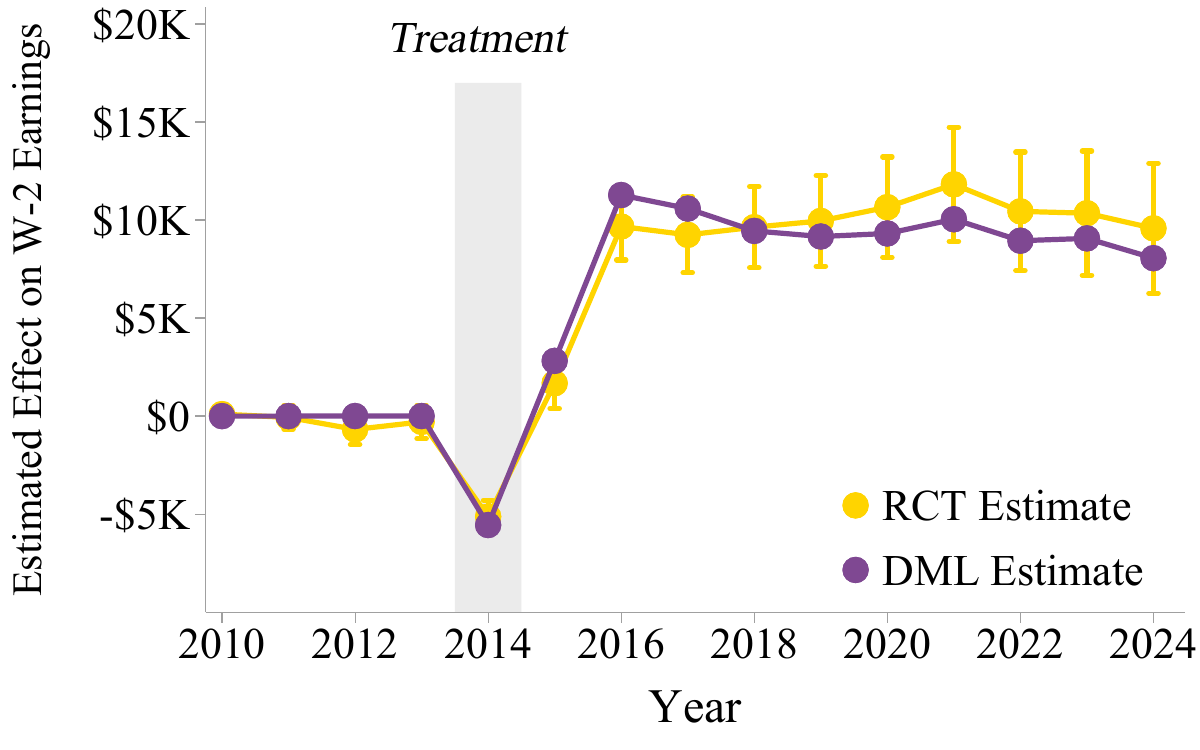}
\end{subfigure}

\vspace{0.35em}

\begin{subfigure}[t]{0.49\linewidth}
\centering
\validationexamplesubcaption{Per Scholas: Earnings Levels}
\includegraphics[width=\linewidth]{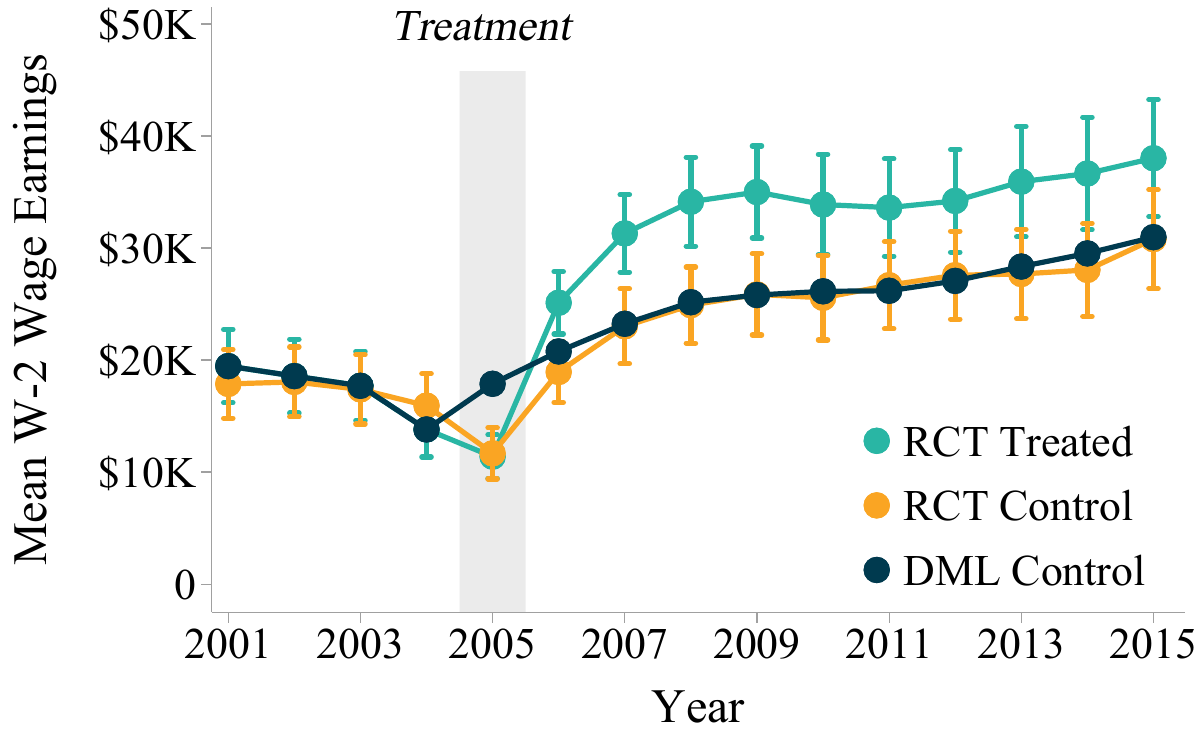}
\end{subfigure}
\hfill
\begin{subfigure}[t]{0.49\linewidth}
\centering
\validationexamplesubcaption{LA Paid Work Experience: Earnings Levels}
\includegraphics[width=\linewidth]{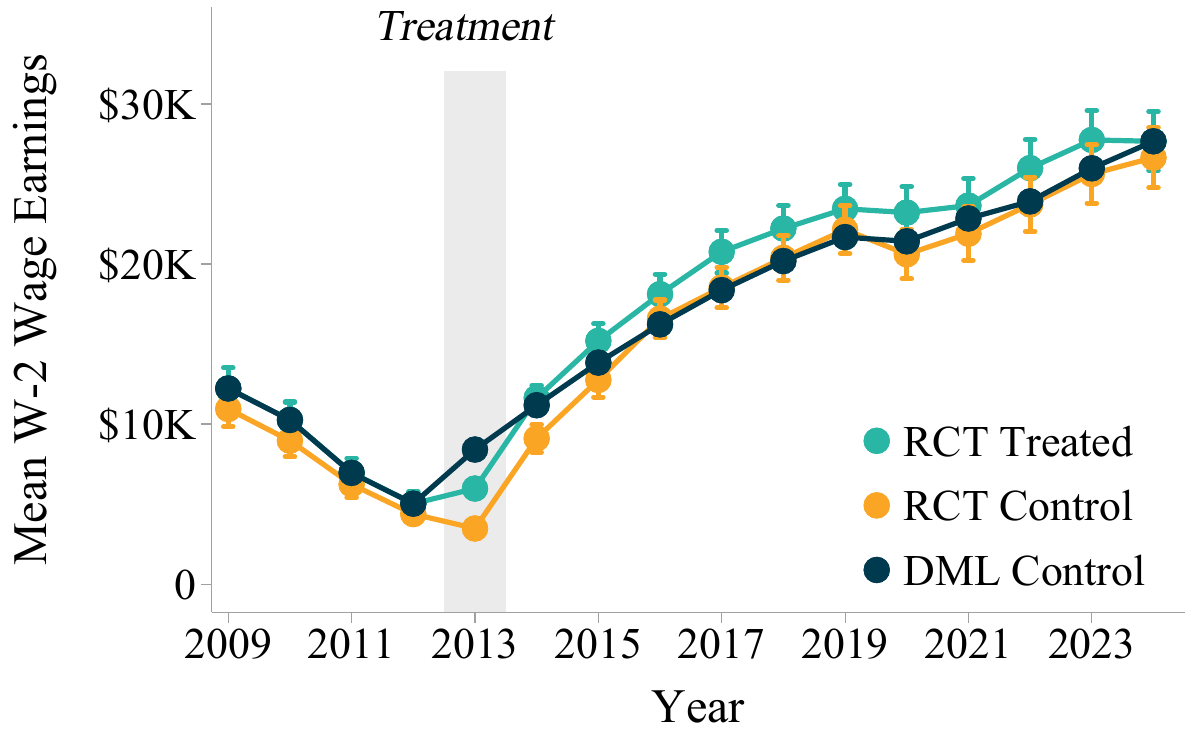}
\end{subfigure}
\fignotes{
These figures compare estimates of the earnings impacts of three workforce programs based on randomized controlled trial (RCT) and observational machine learning (DML) estimators. Panels A, C, and D plot mean annual W-2 earnings for the treatment group, the RCT control group, and the baseline DML estimate of treated participants' counterfactual earnings. Panel A presents annual earnings from 2010 through 2024 for the Year Up United PACE evaluation, pooled across nine sites; Panel C presents annual earnings from 2001 through 2015 for the Per Scholas site in the Sectoral Employment Impact Study; and Panel D presents annual earnings from 2009 through 2024 for the Los Angeles Paid Work Experience program in the Subsidized and Transitional Employment Demonstration. Panels A and B pool PACE's staggered 2013--2014 enrollment years in event time and display event year 0 as 2014. Panel B plots the Year Up United treatment effect estimates obtained by subtracting the RCT and DML counterfactual means, respectively, from the treatment group mean. Vertical bars denote 95\% confidence intervals for the treatment group and RCT control group means in Panels A, C, and D and for the RCT treatment effect estimates in Panel B.
In this and all subsequent figures, dollar amounts are reported in \DollarYear{} dollars.
}
\end{minipage}
\end{figure}
 \clearpage
\begin{figure}[p]
\centering
\captionsetup{labelsep=newline}
\caption{\hspace{0.5em}Randomized Trial vs. Machine Learning Estimates of Workforce Training Program Treatment Effects}
\label{fig:validation-scatter}
\begin{minipage}{\linewidth}
\centering
\includegraphics[width=\linewidth]{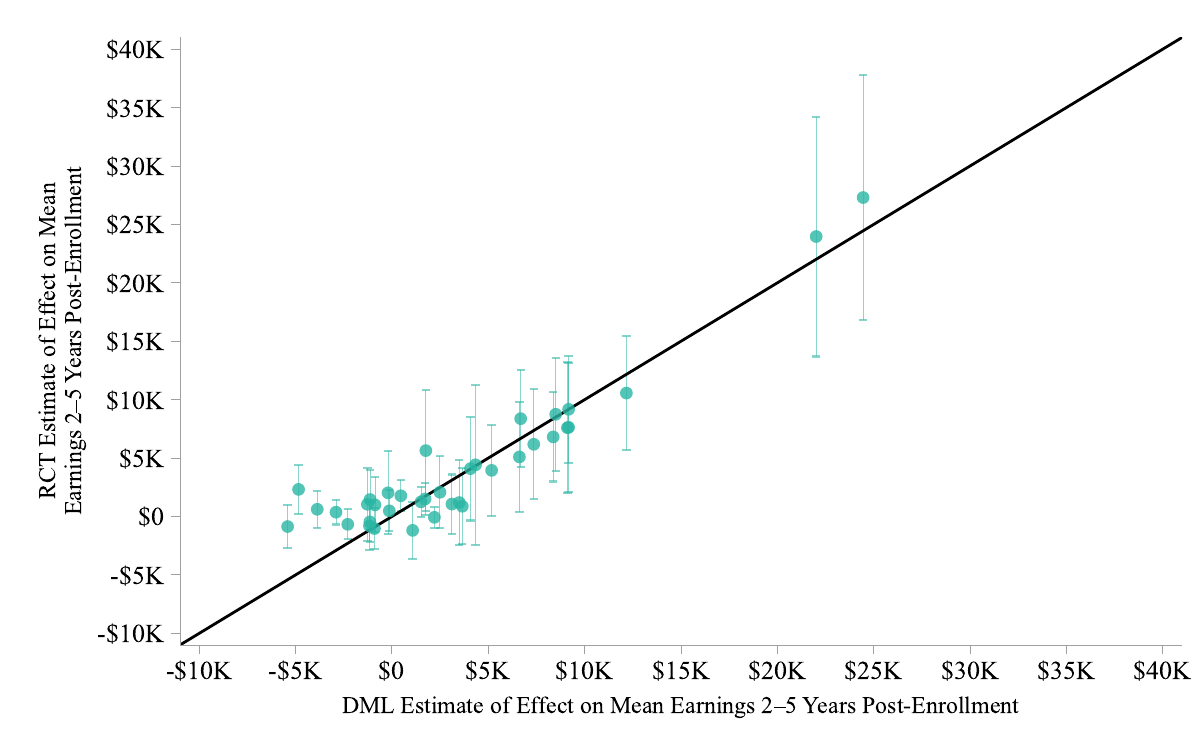}
\fignotes{This figure plots RCT estimates of treatment effects on mean annual W-2 earnings 2--5 years after enrollment in a job training program vs. the corresponding observational machine learning (DML) estimate. The DML estimates are constructed using data from the corresponding treatment group combined with observational data from the population observed in income tax records. The black line is the 45-degree line, where the DML and RCT estimates are equal. Vertical bars depict 95\% confidence intervals for the RCT estimates.
}
\end{minipage}
\end{figure}
 \clearpage
\begin{figure}[p]
\centering
\captionsetup{labelsep=newline}
\caption{\hspace{0.5em}Root Mean Squared Error by Covariate Specification}
\label{fig:covariate_variation}
\begin{minipage}{\linewidth}
\centering
\includegraphics[width=\linewidth]{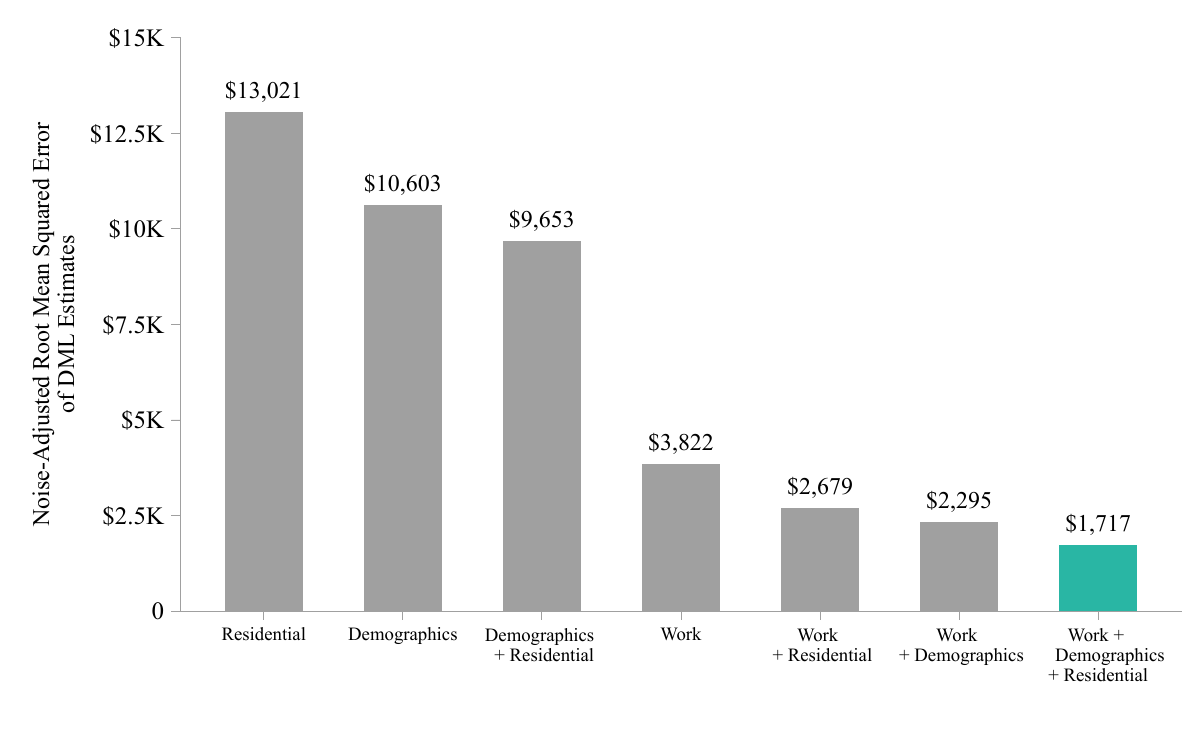}
\fignotes{This figure reports the noise-adjusted RMSE for DML estimates constructed using different sets of pre-program covariates. The noise-adjusted RMSE is calculated as described in Table~\ref{tab:validation-summary}, using the same set of randomized trials and the same outcome. Work refers to work history covariates: total annual W-2 earnings in event years $-5$ through $-1$; indicators for employment (positive W-2 earnings) in event years $-5$ through $-1$; mean earnings and the change in earnings over event years $-3$ through $-1$; and employer industry (2- and 4-digit NAICS codes) in event years $-2$ and $-1$. Demographics refers to demographic covariates: sex; age in event year 0; marital status in event years $-3$ through $-1$; presence of children in event years $-3$ through $-1$. Residential refers to residential history covariates: state and county of residence in event year $-2$, together with characteristics of the residential Census tract in that year: poverty, unemployment, and high school graduation rates in 1990, 2000, and 2010; racial and ethnic composition in 2000; 3rd-grade reading and math achievement in 2015; and district-level standardized achievement measures from the Stanford Education Data Archive. The final (green) bar reports the RMSE of our baseline digital twins estimator, which uses the full set of work, demographic, and residential history covariates.}
\end{minipage}
\end{figure} \clearpage
\begin{figure}[p]
\centering
\captionsetup{labelsep=newline}
\caption{\hspace{0.5em}Root Mean Squared Error by Model Complexity and Sample Size}
\label{fig:model_complexity}
\begin{minipage}{\linewidth}
\centering
\includegraphics[width=\linewidth]{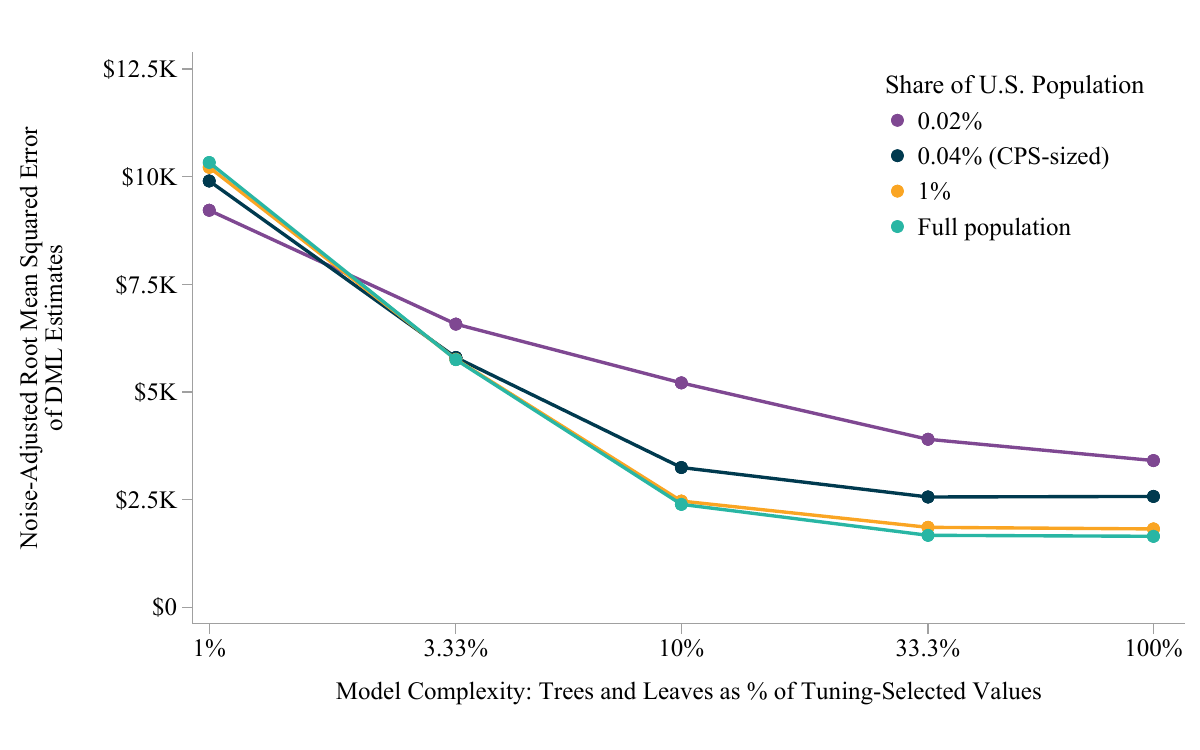}
\fignotes{This figure reports the noise-adjusted RMSE for DML estimates constructed with various comparison sample sizes and model-complexity parameters. The horizontal axis scales the number of trees and leaves to 1, 3.33, 10, 33.3, or 100\% of the tuning-selected values from the baseline DML model and is shown on a logarithmic scale. For each series, we randomly sample 0.02, 0.04, 1, or 100\% of the full U.S. population before applying the same baseline age, wage, and geographic eligibility restrictions described in Section~\ref{empirical-implementation}. The 0.04\% sample is comparable in size to the 1976 CPS sample used by LaLonde, which contained about 136,000 individuals. The noise-adjusted RMSE is calculated using the same set of randomized trials and the same outcome as in Table~\ref{tab:validation-summary}.}
\end{minipage}
\end{figure}

\clearpage
\begin{figure}[p]
\centering
\captionsetup{labelsep=newline}
\captionsetup[subfigure]{font={small,singlespacing},justification=centering}
\caption{\hspace{0.5em}Year Up United Treatment Effects by Year of Enrollment and Site}
\label{fig:yearup-heterogeneity}
\begin{minipage}{\linewidth}
\centering
\begin{subfigure}[t]{\linewidth}
\centering
\caption{Average Treatment Effects by Year of Enrollment}
\includegraphics[width=\linewidth,trim=0 13pt 0 13pt,clip]{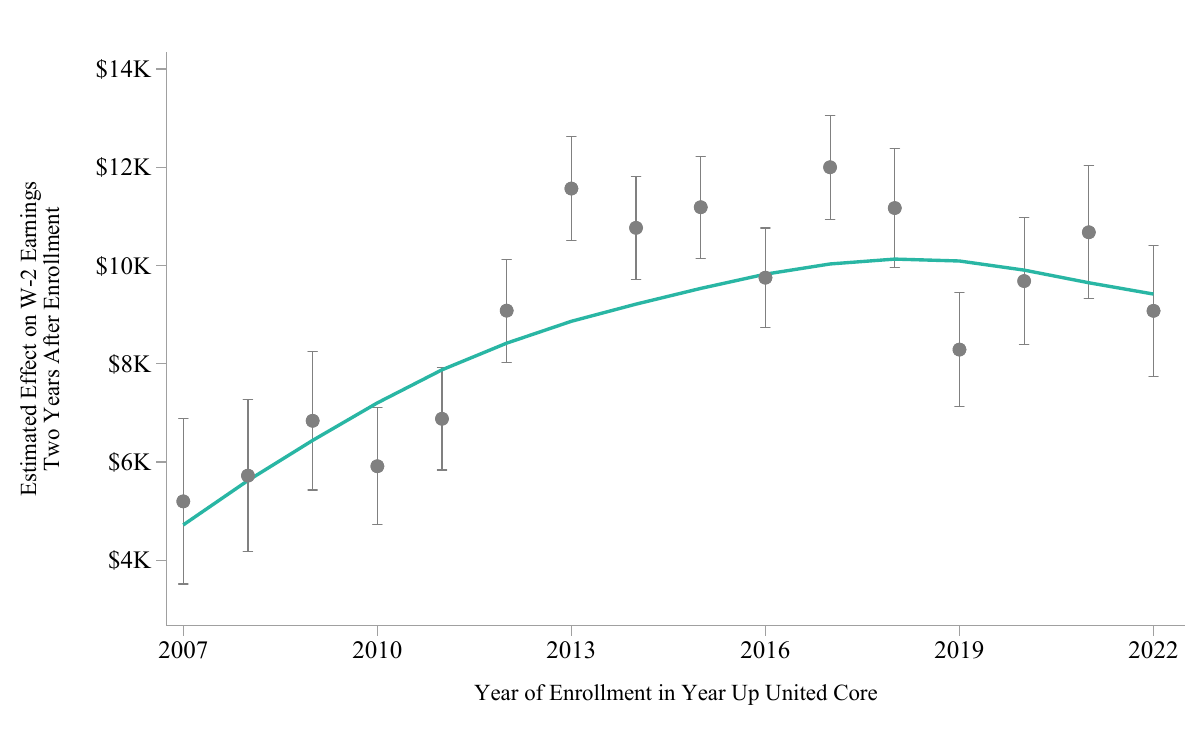}
\end{subfigure}

\vspace{0.25em}

\begin{subfigure}[t]{\linewidth}
\centering
\caption{Average Treatment Effects for Selected Sites by Year of Enrollment}
\includegraphics[width=\linewidth,trim=0 13pt 0 13pt,clip]{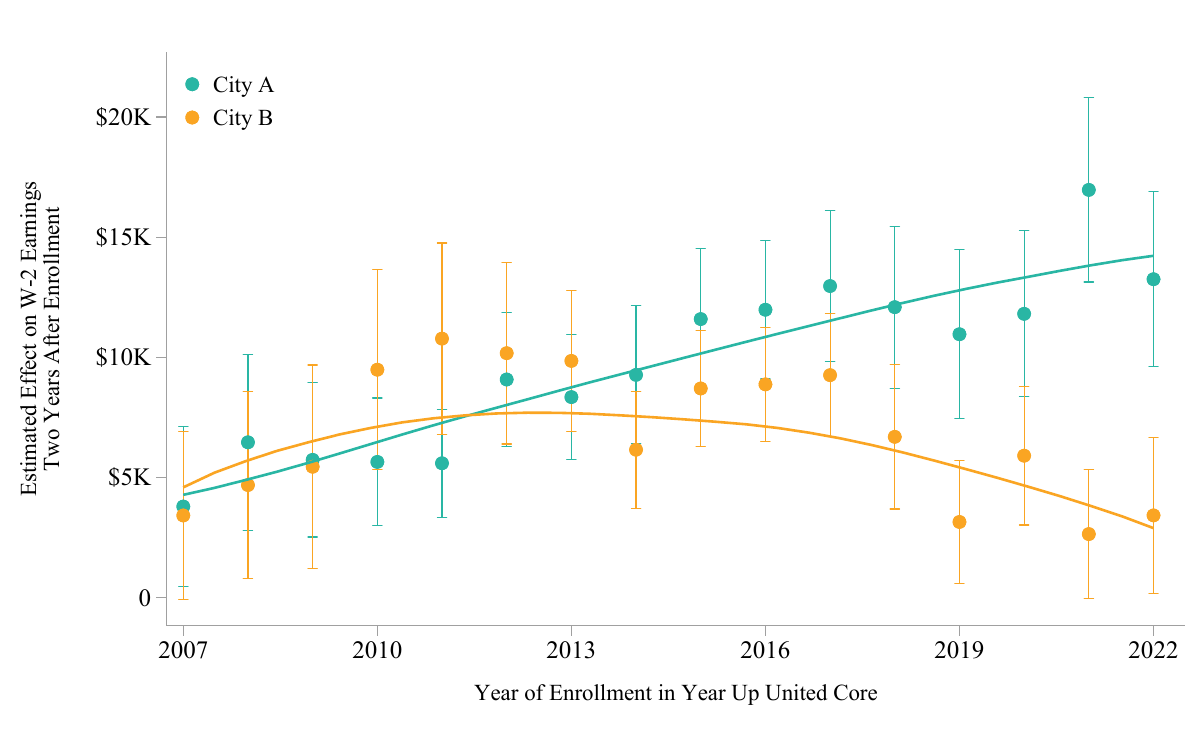}
\end{subfigure}
\fignotes{These figures report baseline digital twins estimates of treatment effects for Year Up United's Core program by enrollment year and site. Participants are assigned to enrollment years based on the calendar year in which they enrolled in Year Up United. The outcome is mean annual W-2 earnings in event year 2. Panel A plots the treatment effect by enrollment year for sites with available estimates, for each enrollment year from \YearUpObsFirstCohort{} through \YearUpLateCohortYear{}. Panel B plots treatment effects by enrollment year for two illustrative sites, anonymized as City A and City B. Solid lines are locally weighted regressions using the points displayed in each panel. Vertical bars denote pointwise 95\% confidence intervals based on sampling error.}

\end{minipage}
\end{figure}

\clearpage
\begin{figure}[p]
\centering
\captionsetup{labelsep=newline}
\caption{\hspace{0.5em}Year Up United Earnings Impacts vs. Target-Sector Placement Rates}
\label{fig:yearup-placement}
\begin{minipage}{\linewidth}
\centering
\includegraphics[width=\linewidth]{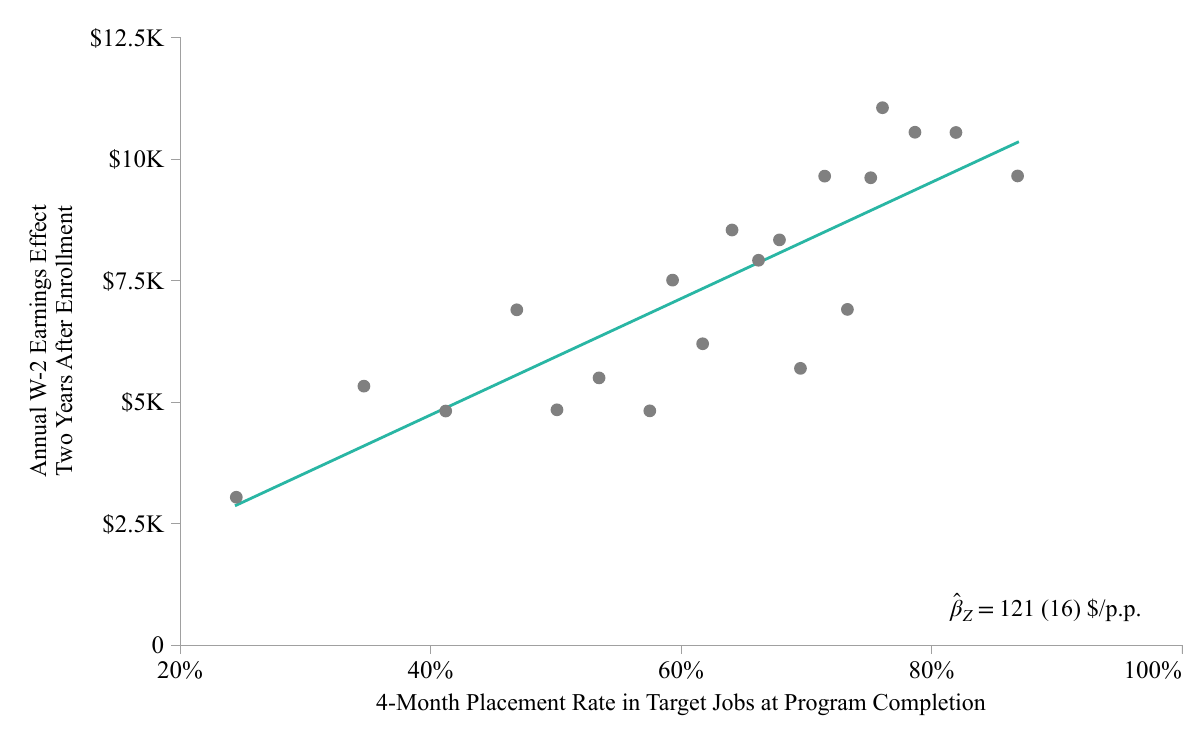}
\fignotes{This figure plots DML earnings impacts against target-sector placement rates for \NYearUpPlacementCells{} site-by-enrollment-year cells in Year Up United's Core and PTC programs. The outcome is mean annual W-2 earnings two years after enrollment. Placement rates measure the share of participants placed into full-time jobs in the target sector within four months of program completion, with noncompleters counted as not placed. Cells are grouped into 20 bins containing approximately equal numbers of cells, and each point reports the participant-weighted mean earnings impact and placement rate within one bin. The solid line is an unweighted least-squares fit through the bin means. The reported coefficient $\hat{\beta}_Z$ is the slope from an unweighted regression of cell-level earnings impacts on placement rates across all \NYearUpPlacementCells{} cells, expressed in dollars per percentage point (p.p.), with standard error reported in parentheses.}
\end{minipage}
\end{figure} \clearpage
\begin{figure}[p]
\centering
\captionsetup{labelsep=newline}
\caption{\hspace{0.5em}Earnings Trajectories for Year Up United Participants Who Switch Jobs vs. Their Digital Twins}
\label{fig:job-industry-switchers}
\begin{minipage}{\linewidth}
\centering
\includegraphics[width=\linewidth]{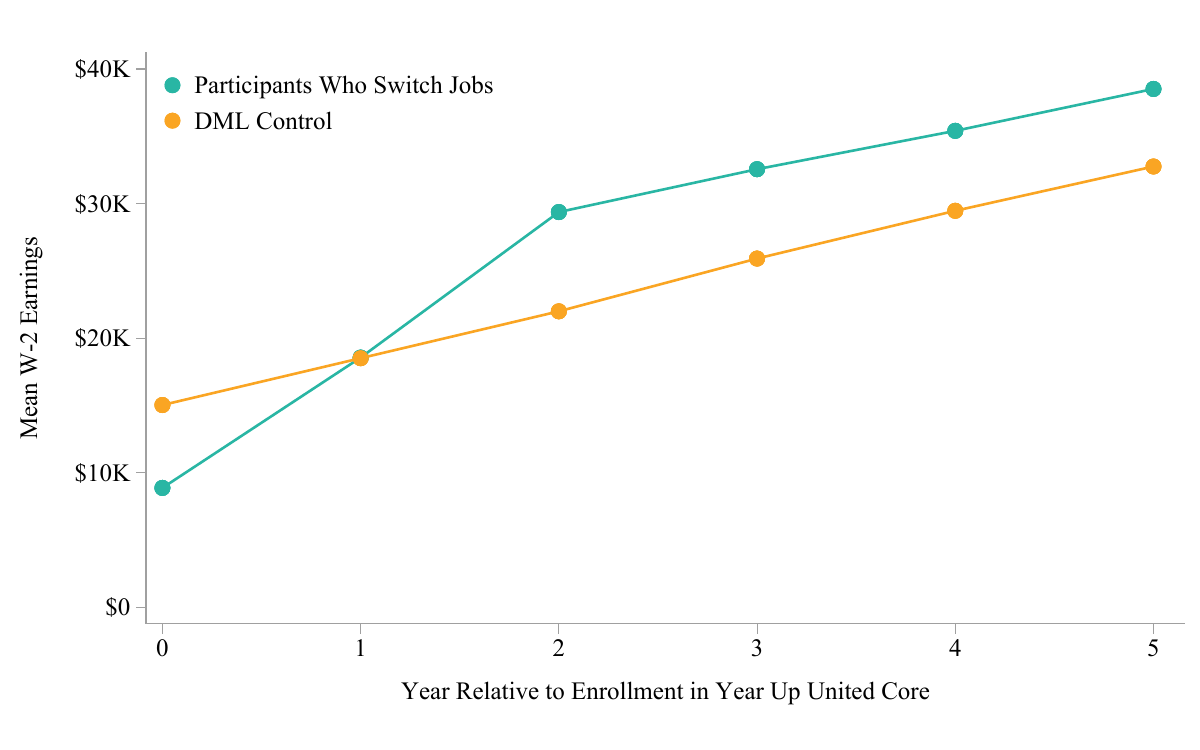}
\fignotes{This figure plots annual W-2 earnings and the digital twins estimate of counterfactual earnings (constructed using the baseline digital twins estimator described in Section~\ref{estimand-and-estimation}) for participants in Year Up United's Core program in enrollment years 2007--2019 who ever switched jobs in the five years after program enrollment. Job switching is defined as an indicator for being associated with a different EIN at any point in event years 2--5 compared to the EIN an individual was associated with in event year 1. \JobSwitcherSharePct{}\% of Year Up United Core participants in enrollment years 2007--2019 switch jobs from event year 1 at some point in event years 2--5.}
\end{minipage}
\end{figure}
 
\begin{table}[p]
  \centering
  \captionsetup{labelsep=newline}
  \caption{\hspace{0.5em}Variance Components of DML and RCT Treatment Effect Estimates}
  \label{tab:validation-summary}

  \setlength{\tabcolsep}{4pt}
  \newcommand{\validationrowstrut}{\rule[-10pt]{0pt}{32pt}}
  \begin{threeparttable}
  \small
  \begin{tabular}{cllcc}
    \toprule
    & Measure & Estimator & Estimate & Std. Error \\
    \midrule

    (1)\validationrowstrut & Raw RMS DML--RCT difference
        & $\left[\frac{1}{J}\sum_{j\in V}
          \left(\widehat{\tau}^{\mathrm{DML}}_j-
          \widehat{\tau}^{\mathrm{RCT}}_j\right)^2\right]^{1/2}$
        & \ValidationRawRMSEDollars{} & \ValidationRawRMSEBootstrapSEDollars{} \\

    (2)\validationrowstrut & Root mean sampling variance of RCT control mean
        & $\left[\frac{1}{J}\sum_{j\in V}
          \widehat{\operatorname{Var}}(\bar Y^{\mathrm{RCT}}_{0j})
          \right]^{1/2}$
        & \ValidationControlNoiseDollars{} & \ValidationControlNoiseBootstrapSEDollars{} \\

    (3)\validationrowstrut & Root mean sampling variance of DML counterfactual
        & $\left[\frac{1}{J}\sum_{j\in V}
          \widehat{\operatorname{Var}}(\widehat\mu^{\mathrm{DML}}_{0j})
          \right]^{1/2}$
        & \PreliminaryValidationDMLCounterfactualNoiseDollars{} & \ValidationDMLCounterfactualNoiseBootstrapSEDollars{} \\

    (4)\validationrowstrut & Noise-adjusted RMSE of DML estimator ($\widehat{\phi}$)
        & $\left[(1)^2-(2)^2-(3)^2\right]^{1/2}$
        & \ValidationNoiseAdjustedRMSEDollars{} & \ValidationNoiseAdjustedRMSEBootstrapSEDollars{} \\

    (5)\validationrowstrut & Root mean sampling variance of DML estimates
        & $\left[
            \frac{1}{J}\sum_{j\in V}
            \widehat{\operatorname{Var}}
            \left(\widehat{\tau}^{\mathrm{DML}}_j\right)
          \right]^{1/2}$
        & \ValidationDMLRMSSEDollars{} & \ValidationDMLRMSSEBootstrapSEDollars{} \\

    (6)\validationrowstrut & RMSE of DML estimates at RCT sample size
        & $\left[(4)^2+(5)^2\right]^{1/2}$
        & \UnscaledValidationDMLTotalRMSEDollars{} & \UnscaledValidationDMLTotalRMSEBootstrapSEDollars{} \\

    (7)\validationrowstrut & RMSE of DML estimates at avg. obs. sample size
        & $\left[(4)^2 + \frac{\bar{N}_1^{\mathrm{RCT}}}{\bar{N}_1^{\mathrm{Obs}}}(5)^2\right]^{1/2}$
        & \ValidationDMLTotalRMSEDollars{} & \ValidationDMLTotalRMSEBootstrapSEDollars{} \\

    (8)\validationrowstrut & RMSE of RCT estimates ($\widehat{\sigma}_{\mathrm{RCT}}$)
        & $\left[
            \frac{1}{J}\sum_{j\in V}
            \widehat{\operatorname{Var}}
            (\widehat{\tau}^{\mathrm{RCT}}_j)
          \right]^{1/2}$
        & \ValidationRMSRCTSEDollars{} & \ValidationRMSRCTSEBootstrapSEDollars{} \\

    (9)\validationrowstrut & Noise-adjusted SD of treatment effects ($\widehat{\sigma}_{\tau}$)
        & $\left[
            \frac{1}{J}\sum_{j\in V}
            \left(\widehat{\tau}^{\mathrm{RCT}}_j-
            \frac{1}{J}\sum_{\ell\in V}\widehat{\tau}^{\mathrm{RCT}}_\ell\right)^2
            -\sigma_{\mathrm{RCT}}^2
          \right]^{1/2}$
        & \ProgramEffectSignalSDDollars{} & \ProgramEffectSignalSDBootstrapSEDollars{} \\

    \addlinespace[6pt]
    \addlinespace[6pt]

    \bottomrule
  \end{tabular}

 \begin{tablenotes}[flushleft]
  \scriptsize\justifying
    \item\textit{Notes: }This table reports estimates of key variance components of our randomized controlled trial (RCT) and baseline double/debiased machine learning (DML) treatment effect estimates. The RCT estimates from \NValidationBenchmarks{} randomized trials are used to validate the observational estimates. The DML estimates are constructed using data from the corresponding treatment group combined with observational data from the population observed in income tax records. The outcome is mean annual W-2 earnings 2--5 years after random assignment, expressed in real \DollarYear{} dollars. The first four rows report components that we use to identify the noise-adjusted RMSE of the DML estimator. $\widehat B_j$ is the difference between the DML and RCT treatment effect estimates in cell $j$. Because the two estimates use the same treated group, their difference depends only on the RCT and DML counterfactuals. The raw dispersion in row 1 therefore reflects both DML bias and sampling error in the two counterfactual estimates. Rows 2 and 3 report these sampling-error components. Subtracting their variances from the squared raw RMSE yields the noise-adjusted RMSE in row 4, denoted $\widehat{\phi}$. Rows 5 through 7 incorporate sampling variation in the treatment effect estimates themselves. The total RMSE of a DML estimate combines its RMS bias from row 4 with its treatment-effect sampling variance. Row 6 reports this total RMSE at the RCT sample sizes, while row 7 reports it at the average observational sample size. To calculate row 7, we multiply the average DML sampling variance by $\bar{N}_1^{\mathrm{RCT}} / \bar{N}_1^{\mathrm{Obs}}$, where $\bar{N}_1^{\mathrm{RCT}}$ is the mean number of treated individuals across validation cells and $\bar{N}_1^{\mathrm{Obs}}$ is the mean number of treated individuals across observational site-by-cohort cells for Year Up United, Project QUEST, and Per Scholas---the programs for which we have both experimental and observational data in our sample. RCT treatment effect estimates are unbiased under random assignment, so their total RMSE is given by their sampling error alone. Row 9 reports the standard deviation of RCT treatment effect estimates across the \NValidationBenchmarks{} randomized trials after subtracting their average sampling variance. Standard errors are estimated using \NValidationBootstrapReplications{} program-level cluster bootstrap replications, with the \NValidationProgramModels{} distinct programs described in Section~\ref{program-records-vcodex} serving as clusters. In each replication, we sample programs with replacement, include all trials associated with each sampled program, and recalculate rows 1--9 giving each trial equal weight. Approximately 5.9\% of bootstrap replications in row 4 and 0.6\% of replications in row 9 produce a negative overall noise-adjusted variance estimate; these estimates are set to zero before taking the square root.
  \end{tablenotes}
  \end{threeparttable}
\end{table}
 \clearpage
\begin{table}[p]
  \centering
  \captionsetup{labelsep=newline}
  \caption{\hspace{0.5em}Noise-Adjusted RMSE by Covariates, Model, and Comparison Sample}
  \label{tab:estimator-comparison}

  \small
  \begin{minipage}{0.78\linewidth}
  \begin{threeparttable}
  \begin{tabular}{@{}p{0.34\linewidth}@{}
    >{\centering\arraybackslash}p{0.15\linewidth}@{}
    >{\centering\arraybackslash}p{0.15\linewidth}@{}p{0.06\linewidth}@{}
    >{\centering\arraybackslash}p{0.15\linewidth}@{}
    >{\centering\arraybackslash}p{0.15\linewidth}@{}}
    \toprule
    \textit{Model}
      & \multicolumn{2}{c}{Parametric (OLS)}
      & & \multicolumn{2}{c}{Flexible (DML)} \\
    \cmidrule(lr){2-3}\cmidrule(lr){5-6}
    \begin{tabular}[c]{@{}l@{}}\textit{Comparison sample}\end{tabular}
      & \begin{tabular}[c]{@{}c@{}}CPS-sized\\[-2pt]sample\\[-1pt]{\scriptsize\itshape LaLonde-style}\\[-3pt]{\scriptsize\itshape restrictions}\end{tabular}
      & \begin{tabular}[c]{@{}c@{}}Population-scale\\[-2pt]sample\\[-1pt]{\scriptsize\itshape Targeted age and}\\[-3pt]{\scriptsize\itshape geographic restrictions}\end{tabular}
      & & \begin{tabular}[c]{@{}c@{}}CPS-sized\\[-2pt]sample\\[-1pt]{\scriptsize\itshape LaLonde-style}\\[-3pt]{\scriptsize\itshape restrictions}\end{tabular}
      & \begin{tabular}[c]{@{}c@{}}Population-scale\\[-2pt]sample\\[-1pt]{\scriptsize\itshape Targeted age and}\\[-3pt]{\scriptsize\itshape geographic restrictions}\end{tabular} \\
    \addlinespace[1pt]
     & (1) & (2) & & (3) & (4) \\
    \midrule
    \rule[-5pt]{0pt}{19pt}LaLonde covariates & \$\ComparisonOLSSmallLaLondeRMSE{} & \$\ComparisonOLSLargeLaLondeRMSE{} & & \$\ComparisonDMLSmallLaLondeRMSE{} & \$\ComparisonDMLLargeLaLondeRMSE{} \\
    \rule[-5pt]{0pt}{19pt}Dehejia--Wahba covariates & \$\ComparisonOLSSmallDehejiaWahbaRMSE{} & \$\ComparisonOLSLargeDehejiaWahbaRMSE{} & & \$\ComparisonDMLSmallDehejiaWahbaRMSE{} & \$\ComparisonDMLLargeDehejiaWahbaRMSE{} \\
    \rule[-5pt]{0pt}{19pt}Expanded covariates & \$\ComparisonOLSSmallExpandedRMSE{} & \$\ComparisonOLSLargeExpandedRMSE{} & & \$\ComparisonDMLSmallExpandedRMSE{} & \$\ComparisonDMLLargeExpandedRMSE{} \\
    \bottomrule
  \end{tabular}
  \begin{tablenotes}[flushleft]
    \scriptsize\justifying
    \item\textit{Notes: }This table reports noise-adjusted RMSE across \NValidationBenchmarks{} equally weighted randomized trials for mean annual W-2 earnings 2--5 years after random assignment, in real \DollarYear{} dollars. The specifications build from a replication of \citet{lalonde1986evaluating} in the top-left cell to our baseline DML specification in the bottom-right cell, varying three dimensions of the estimation data and model: covariate richness, comparison sample, and model flexibility. LaLonde covariates include prior-year earnings and employment, demographic characteristics, and metropolitan-area residence, as described in Section~\ref{performance-drivers}. Dehejia--Wahba covariates add second-lag W-2 earnings to that set. We compare these sets with the richer covariates used in our baseline specification: five years of earnings and employment history, employer industry, demographics, and residential characteristics, as defined in Section~\ref{empirical-implementation}. The comparison samples differ in both size and eligibility restrictions. To replicate LaLonde's sample, we randomly draw 136,000 individuals---the approximate size of the 1976 CPS---and apply eligibility restrictions in the spirit of his analysis: age below 55 and prior-year W-2 earnings below \$100,000. Results for the CPS-sized sample are averaged across 50 repetitions for stability. The population-scale sample is constructed using a 10\% random draw from the full U.S. population, subject to the age, wage, and geographic eligibility restrictions used in our baseline estimator and described in Section~\ref{empirical-implementation}. We calculate noise-adjusted RMSE as described in Table~\ref{tab:validation-summary}, subtracting sampling variance in the RCT control mean and the observational estimator's counterfactual mean. For OLS, we estimate the counterfactual mean's sampling variance using the difference between the treated-mean and treatment-effect influence functions; the latter is obtained via Frisch--Waugh--Lovell.
  \end{tablenotes}
  \end{threeparttable}
  \end{minipage}
\end{table}
 
\newpage

\begin{appendices}
\section{Coverage Intervals and Treatment Effect Comparisons}\label{app:rmse-coverage}

This appendix establishes two results. First, it establishes a 95\% coverage result for a given DML estimate based on its RMSE. Second, it shows that p-values for comparisons between DML estimates across cells constructed using these confidence intervals are conservative when biases across cells are positively correlated.

\textit{95\% Coverage Interval.} We establish the RMSE-based coverage interval used in Section~\ref{validation-design} by adapting the argument in \citet[Appendix~A.2]{bernard2025trust}. In addition to Assumption~\ref{assumption:common-marginal-bias}, suppose that the marginal bias distribution is normal, $B_k\sim\mathcal{N}(\mu_B,\nu_B^2)$. By $\sqrt{n}$-inference for the estimator, the target estimator's sampling error satisfies
\[
e_k
=
\widehat{\tau}_k-\tau_k-B_k,
\qquad
\frac{e_k}{\widehat{\sigma}_k}
\xrightarrow{d}
\mathcal{N}(0,1).
\]
Following the decomposition in \citet[Appendix~A.2]{bernard2025trust},
\[
\widehat{\tau}_k-\tau_k
=
\left(\widehat{\tau}_k-\tau_k-B_k\right)+B_k
=
e_k+B_k
\]
Suppose that $e_k$ is independent of $B_k$. Then as the number of validation cells $J$ and the target-cell sample size $n_k$ grow, $\widehat{\phi}^{\,2} \, \xrightarrow{p} \, \mathbb{E}[B_k^2]$, \,$e_k \, \xrightarrow{p} \, 0$, and
\begin{align*}
    \widehat{\tau}_k - \tau_k \xrightarrow{d} \mathcal{N}(\mu_B, \nu_B^2).
\end{align*}
Meanwhile, squared estimated RMSE converges in probability to $E[B_k^2] = \mu_B^2 + \nu_B^2$. By Slutsky's theorem, the limiting coverage of
\[
\widehat{\tau}_k
\ \pm\
1.96\left(
\widehat{\phi}^{\,2}
+\widehat{\sigma}_k^2
\right)^{1/2}
\]
is equal to that of the interval $\widehat{\tau}_k \pm 1.96 \, E[B_k^2]^{1/2}$, which is asymptotically conservative relative to the correct 95\% confidence interval $\widehat{\tau}_k \pm 1.96 \, \nu_B$ whenever $\mu_B \neq 0$ and achieves the minimum 95\% coverage when $\mu_B = 0$. Thus, averaging over target-cell bias and within-cell sampling error, the interval contains the target cell's true treatment effect with probability at least 95\%. 

Similarly, using the RMSE of a typical validation cell yields a coverage rate of at least 95\% on average across validation cells as the size and number of validation cells increase. Because the biases $B_j$ across validation cells are draws from the same normal distribution, the argument above applies with the validation cell $j$ in place of the target cell $k$ as the number of validation cells increases. The generic interval $\widehat{\tau}_j \pm 1.96 \; \text{RMSE}(\widehat{\tau})$ has asymptotic coverage of at least 95\% for a validation cell $j$ averaging over the distribution of bias, since $\text{RMSE}(\widehat{\tau})^2 = \widehat{\phi}^2 + \frac{1}{J} \sum_{j \in V} \widehat{\sigma}_j^2 \xrightarrow{p} E[B_j^2]$. Hence, the average asymptotic coverage rate across validation cells is at least 95\%. $\square$

\textit{P-Values for Treatment Effect Comparisons.} Under similar assumptions, we consider two sites $j$ and $k$ for which treatment effect estimates $\widehat{\tau}_j$ and $\widehat{\tau}_k$ are available and establish that confidence intervals for treatment effect differences across cells in Section~\ref{applications} are asymptotically conservative when the bias across sites is positively correlated, $\rho = \text{Corr}(B_k, B_j) > 0$.
\begin{align*}
    (\widehat{\tau}_j - \widehat{\tau}_k) - (\tau_j  - \tau_k) &= (\widehat{\tau}_j - \tau_j - B_j) - (\widehat{\tau}_k - \tau_k - B_k) + B_j - B_k \\
    &= e_j - e_k + B_j - B_k
\end{align*}
$B_k, B_j$ are jointly normal, hence $B_j - B_k \sim \mathcal{N}(0,  2 \nu_B^2 (1-\rho))$. Assume that sampling errors across sites are independent. As the number of sites $J$ increases, $\widehat{\phi}^2 \xrightarrow{p} E[B_j^2]$ and as $n_j, n_k$ increase, $e_j, e_k \xrightarrow{p} 0$.
\begin{align*}
    (\widehat{\tau}_j - \widehat{\tau}_k) - (\tau_j  - \tau_k) \xrightarrow{d} \mathcal{N}(0, 2 \nu_B^2 (1-\rho))
\end{align*}
Under the null $H_0: \tau_j - \tau_k = 0$, the confidence interval
\begin{align*}
    \widehat{\tau}_j - \widehat{\tau}_k \ \pm \ 1.96 \ \Big(2\widehat{\phi}^2 + \widehat{\sigma}_j^2 + \widehat{\sigma}_k^2\Big)^{1/2}
\end{align*}
has a limiting coverage as $J, n_j, n_k \rightarrow \infty$ equal to the interval
\begin{align*}
    \widehat{\tau}_j - \widehat{\tau}_k \ \pm \ 1.96 \ (2 \, E[B_k^2])^{1/2}
\end{align*}
which is asymptotically conservative as long as $2 \nu_B^2 \, (1-\rho) < 2 \, E[B_k^2]$. This is true for any $\mu_B$ as long as $\rho > 0$. Equivalently, the corresponding two-sided tests yield conservative p-values for comparisons between treatment effects. $\square$ \end{appendices}
\clearpage
\begingroup
\setcounter{figure}{0}
\renewcommand{\thefigure}{\arabic{figure}}
\renewcommand{\theHfigure}{appendix.\arabic{figure}}
\captionsetup[figure]{name=Appendix Figure}

\clearpage
\begin{figure}[p]
\centering
\captionsetup{labelsep=newline}
\caption{\hspace{0.5em}RCT vs. Machine Learning Estimates with Confidence Intervals for Both Estimators}
\label{fig:validation-scatter-both-ci}
\begin{minipage}{\linewidth}
\centering
\includegraphics[width=\linewidth]{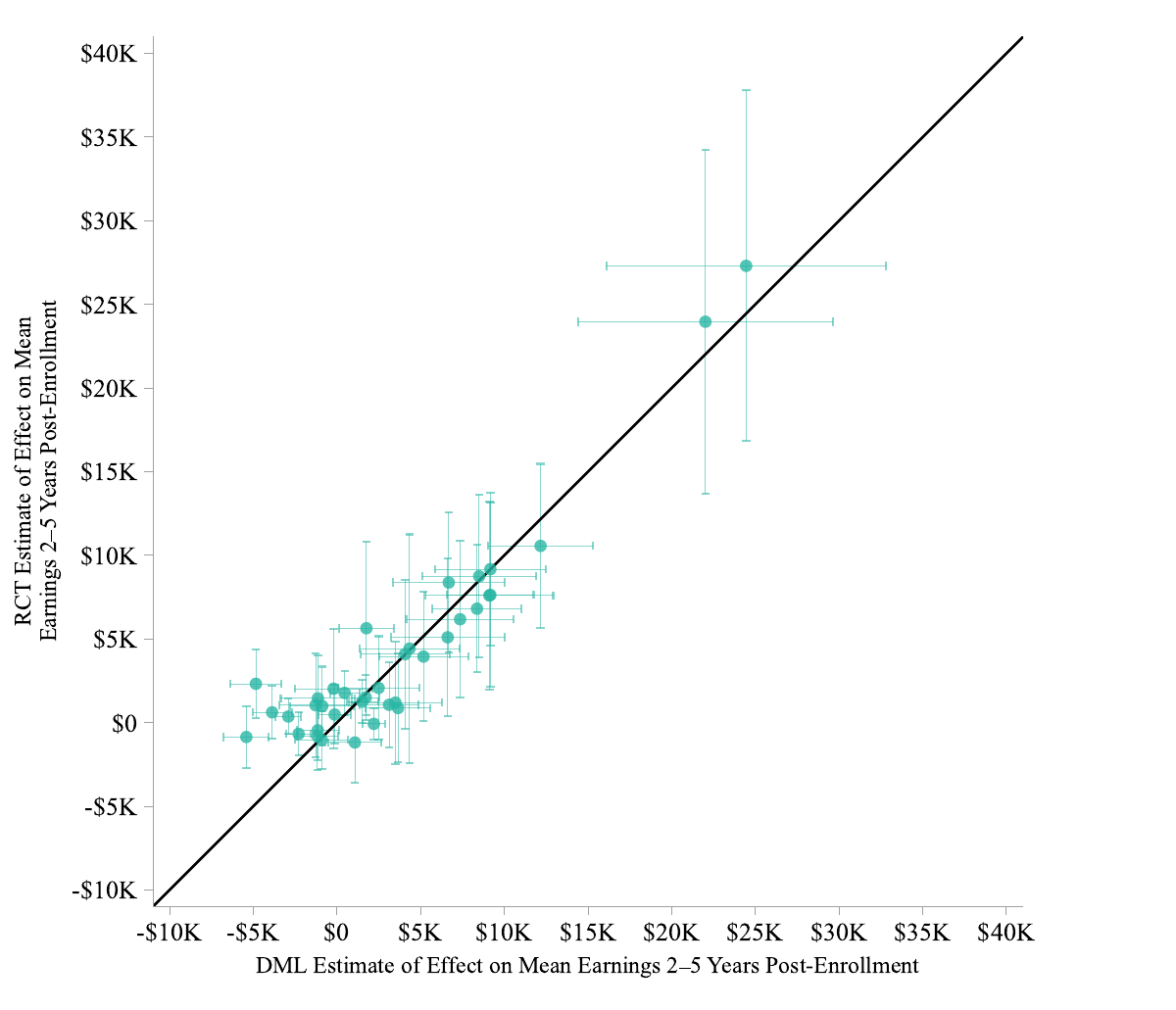}
\fignotes{This figure reproduces Figure~\ref{fig:validation-scatter} with confidence intervals for both estimators. Horizontal bars denote 95\% confidence intervals for the DML estimates, constructed as the point estimate plus or minus 1.96 times its standard error. These intervals account for sampling error but not potential selection bias. Both axes use the same limits and scale.}

\end{minipage}
\end{figure}
 \begin{figure}[!htbp]
\centering
\captionsetup{labelsep=newline}
\caption{\hspace{0.5em}Covariate Importance in Predicting Counterfactual Earnings}
\label{fig:shap-importance}
\begin{minipage}{\linewidth}
\centering
\includegraphics[width=\linewidth]{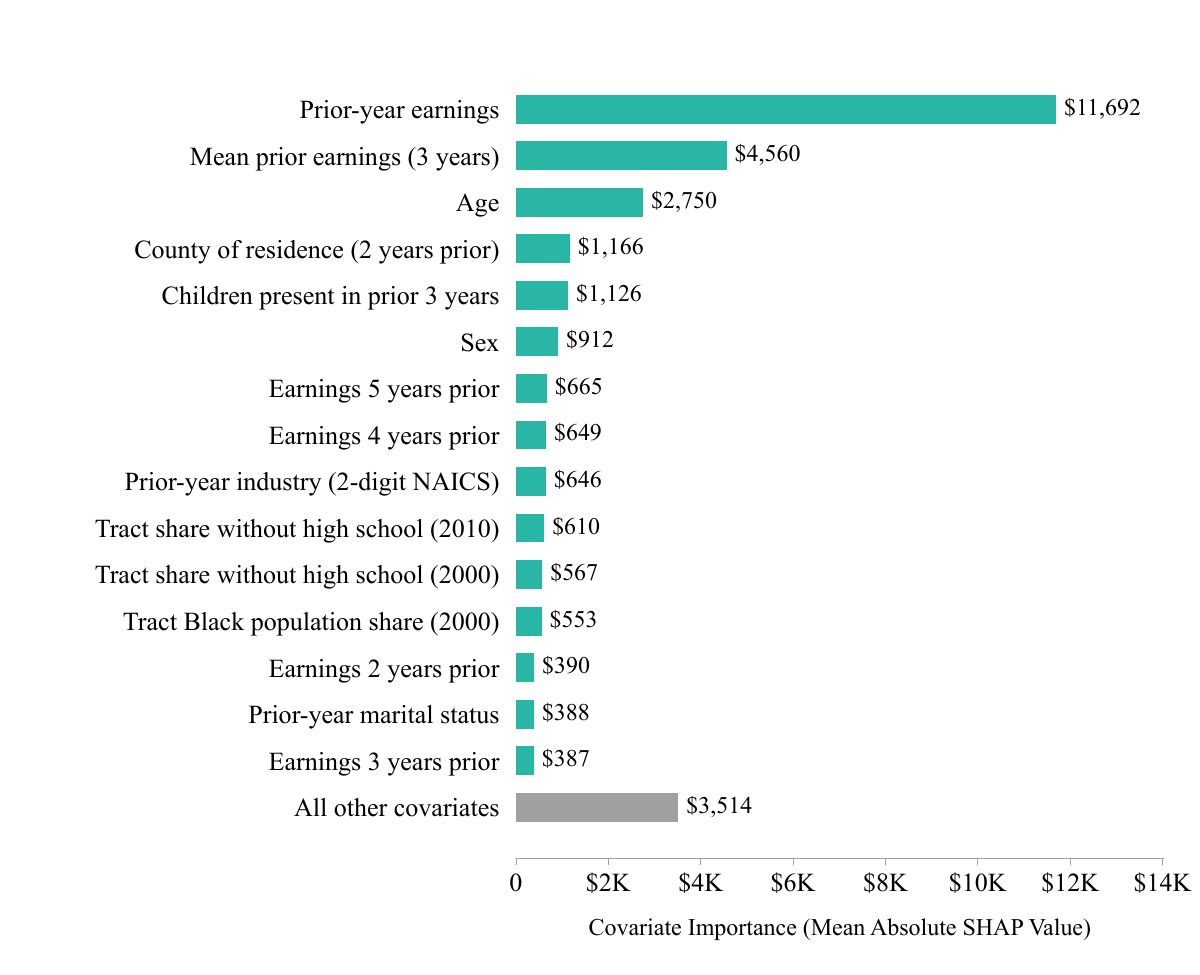}
\fignotes{This figure summarizes covariate importance in the baseline digital twins outcome model by reporting the average magnitude, in dollars, of each covariate's contribution to predicted counterfactual mean annual W-2 earnings 2--5 years after enrollment. Importance is measured using mean absolute SHAP (Shapley additive explanations) values, computed using TreeSHAP for the gradient-boosted decision trees. These values capture the magnitude of each covariate's contribution relative to the model's baseline prediction, irrespective of direction. For each of the \NValidationBenchmarks{} validation cells, we compute SHAP values for all treated individuals using the first-fold outcome model. For each covariate, we average the absolute SHAP values within each cell, then take an equally weighted average across cells. The teal bars show the 15 covariates with the highest importance, ranked in descending order; the grey bar, labeled ``All other covariates,'' sums the importance measures for the remaining covariates.}
\end{minipage}
\end{figure}
 \clearpage
\endgroup
 
\end{document}